\newcommand{\msun}{\mbox{$M_{\odot}$}}

\newcommand{\mmax}{\mbox{$M_{\rm max}$}}
\newcommand{\mmin}{\mbox{$M_{\rm min}$}}

\newcommand{\redchisq}{\mbox{$\chi_{\nu}^{2}$}}
\newcommand{\kms}{\mbox{km$\,$s$^{-1}$}}
\newcommand{\nrextended}{\mbox{305}}

\newcommand{\slopesmc}{\mbox{$-2.0\pm0.2$}}
\newcommand{\slopelmc}{\mbox{$-1.9\pm0.1$}}
\newcommand{\slopengcf}{\mbox{$-2.1\pm0.1$}}

\newcommand{\slopengcsone}{\mbox{$-1.7\pm0.5$}}
\newcommand{\slopengcstwo}{\mbox{$-2.4\pm0.1$}}
\newcommand{\breakngcs}{\mbox{$-8.9\pm0.4$}}
\newcommand{\mstarngcs}{\mbox{$-10.2\pm0.6$}}

\newcommand{\slopemfone}{\mbox{$-2.0\pm0.1$}}
\newcommand{\slopemftwo}{\mbox{$-2.7\pm0.3$}}
\newcommand{\breakmf}{\mbox{$-9.3\pm0.4$}}
\newcommand{\mstarmf}{\mbox{$-10.3\pm0.5$}}

\newcommand{\slopeantone}{\mbox{$-1.9\pm0.1$}}
\newcommand{\slopeanttwo}{\mbox{$-2.8\pm0.4$}}
\newcommand{\breakant}{\mbox{$-10.3\pm0.4$}}
\newcommand{\mstarant}{\mbox{$-11.4\pm0.4$}}

\newcommand{\slopemodone}{\mbox{$-2.0$}}
\newcommand{\slopemodtwo}{\mbox{$-2.5$}}
\newcommand{\breakmod}{\mbox{$-9.3$}}

\newcommand{\dr}{\mbox{${\rm d}$}}

\documentclass[twocolumn]{aa}
\usepackage{graphicx}
\usepackage{natbib}
\bibpunct{(}{)}{;}{a}{}{,} 
\begin{document}
   \title{The luminosity function of young star clusters: implications for the maximum mass and luminosity of clusters}

  \titlerunning{The luminosity function of young star clusters}

   \author{M. Gieles \inst{1} \and S.S. Larsen \inst{2} \and  N. Bastian \inst{1,3} \and I.T. Stein \inst{1}
          } 

   \offprints{gieles@astro.uu.nl}

   \institute{$^1$Astronomical Institute, Utrecht University, 
              Princetonplein 5, NL-3584 CC Utrecht The Netherlands \\
              \email{gieles@astro.uu.nl} \\
	      $^2$European Southern Observatory, ST-ECF, Karl-Schwarzchild-Strasse 2
	       D-85748 Garching bei M\"{u}nchen, Germany \\
	        $^3$Department of Physics and Astronomy, University College London,
              Gower Street, London, WC1E 6BT\\
             }

   \date{Received June 8, 2005; accepted  December 29, 2005}


   \abstract{We introduce a method to relate a possible truncation of
the star cluster mass function at the high mass end to the shape of
the cluster luminosity function (LF). We compare the observed LFs of
five galaxies containing young star clusters with synthetic cluster
population models with varying initial conditions. The LF of the SMC,
the LMC and NGC~5236 are characterized by a power-law behavior $N\,\dr
L \propto L^{-\alpha}\,\dr L$, with a mean exponent of $<\alpha> = 2.0
\pm 0.2$.  This can be explained by a cluster population  formed
with a constant cluster formation rate, in which the maximum cluster
mass per logarithmic age bin is determined by the size-of-sample
effect and therefore increases with log(age/yr).  The LFs of NGC~6946
and M51 are better described by a double power-law distribution or a
Schechter function. When a cluster population has a mass function that
is truncated below the limit given by the size-of-sample effect, the
total LF shows a bend at the magnitude of the maximum mass, with the
age of the oldest cluster in the population, typically a few Gyr due
to disruption.  For NGC~6946 and M51 this  suggests a maximum
mass of $\mmax = 0.5-1\times10^6 \msun$, although the bend is only a 1-2
$\sigma$ detection. Faint-ward of the bend the LF has the same slope
as the underlying initial cluster mass function and bright-ward of the
bend it is steeper. This behavior can be well explained by our
population model. We compare our results with the only other galaxy
for which a bend in the LF has been observed, the ``Antennae''
galaxies (NGC~4038/4039). There the bend occurs brighter than in
NGC~6946 and M51, corresponding to a maximum cluster mass of $\mmax =
1.3-2.5\times10^6\,\msun$. Hence,  if the maximum cluster mass has a
physical limit, then it can vary between different galaxies.
The fact that we only observe this bend in the LF in the ``Antennae''
galaxies, NGC~6946 and M51 is because there are enough clusters
available to reach the limit. In other galaxies there might be a
physical limit as well, but the number of clusters formed or observed
is so low, that the LF is not sampled up to the luminosity of the
bend. The LF can then be approximated with a single power-law
distribution, with an index similar to the initial mass function index.
}

\maketitle

\section{Introduction}
\label{sec:introduction}

The study of young extra-galactic star clusters has become a whole new
field of research since the discovery of young massive clusters. The
{\it Hubble Space Telescope (HST)} has made it possible to resolve
these objects and to undertake systematic studies of the nature of
these objects. Young clusters with masses in the range of our Milky
Way globular clusters ($10^4 - 10^6 \msun$) have been found in merging
galaxies like the ``Antennae'' \citep{1995AJ....109..960W},
interacting galaxies like M51 (\citealt{2000MNRAS.319..893L};
\citealt{2005A&A...431..905B}), starburst galaxies
(\citealt{1995AJ....110.2665M}; \citealt{2003MNRAS.343.1285D}) and
even non-interacting spiral galaxies \citep{1999A&A...345...59L}.

Recently, a relatively young and very massive cluster was discovered
in the merger remnant NGC~7252 \citep{1993AJ....106.1354W}. Its
(dynamical) mass was confirmed to be as high as $\sim 10^8 \msun$
\citep{2004A&A...416..467M}. This suggests that there are clusters
that fill the gap between the mass range of star clusters and that of
dwarf galaxies. However, it remains to be seen if every galaxy is able
to produce such massive cluster or if there are physical limitations
to the maximum mass of star clusters in different environments.

The LF of star clusters is a powerful tool when studying populations
of star clusters. It indirectly gives us information about the
underlying mass function (MF). \citet{1999ApJ...527L..81Z} showed
for the young clusters in the ``Antennae'' galaxies that the cluster
initial mass function (CIMF) can be well approximated by a power-law
distribution: $N(M)\,\dr M\propto\!~M^{-\alpha^{\prime}}\,\dr M$, with
an exponent\footnote{Through the remainder of this work we will use
$\alpha^{\prime}$ for the exponent of the cluster initial mass
function and $\alpha$ for the exponent of the total cluster luminosity
function.} of $-\alpha^{\prime} = -2$. They derived the ages of the
clusters using reddening-free parameters, which is necessary to extract
the initial mass function from the total luminosity function. The
exponent of the CIMF ($-\alpha^{\prime}$) is found to be close to $-2$
in a wide range of galaxy environments down to masses of $\sim 10^3
\msun$ \citep{2003MNRAS.342..259D}. This resembles the mass
function of molecular clouds \citep{1987ApJ...319..730S}, consistent
with clusters forming from molecular clouds. The LF can also be
approximated with a power-law distribution: $N(L)\,\dr
L\propto\!~L^{-\alpha}\,\dr L$, where values for the exponent
$-\alpha$ are found in a range of $-2.4$ up till $-1.7$
(\citealt{2002AJ....124.1393L}, hereafter L02;
\citealt{2003dhst.symp..153W}). In general the indices of the 
bright LFs are smaller (i.e. steeper) than the index of the CIMF
(L02). One of the unanswered questions is whether the difference in
slope between the mass function and LF have a physical meaning or if
it is a result of different measurement techniques and artifacts like
the contamination by stars at lower luminosities.

\citet{1999AJ....118.1551W} observe a
distinct bend in the LF of young star clusters in the ``Antennae''
galaxies, where the slope faint-ward is shallower than the slope
bright-ward of the bend. The exact slopes depend on the different ways
of correcting for stellar contamination. They argue that this could be the
progenitor turn-over of the peak that is observed in the luminosity
function of old globular clusters which appears at $M_V~\simeq~-7.2$,
corresponding to a mass of $2\times10^5
\msun$. The difficulty with directly relating the LF to a CIMF is that
the LF consists of clusters of different ages, and the mass-to-light
ratio of clusters changes drastically when clusters age. Between
$10^7$ and $10^9$ year, a star cluster of constant mass fades about 4
magnitudes in the $V$-band.  Recently, \citet{mengel05} observed
the clusters in the ``Antennae'' galaxies in the Ks-band. They also
find a double power-law behavior in the LF and argue that the mass
function has a turn-over.

When there is no physical limit to the cluster mass or if there are
not enough clusters to sample the CIMF up to any such limit, the mass
of the most massive cluster will be determined by the total number of
clusters and the slope of the CIMF (\citealt{2003AJ....126.1836H},
hereafter H03).  A similar idea was posed by
\citet{2004MNRAS.350.1503W}, who suggest that the maximum cluster mass
in a galaxy is determined by the star formation rate in that galaxy.
It has not been shown yet in a large sample of galaxies that the most
massive cluster is a result of size-of-sample effects. H03 only showed
that this is the case in the SMC and the LMC. In this study we add M51
to the sample in order to see if the most massive object in this
galaxy is also a result of size-of-sample effect or if there is a
physical limit above which clusters cannot form in this galaxy. In
addition, we use a method to relate a possible truncation of the mass
function at the high mass end with the shape of cluster LF. To this
end we introduce an analytical model to generate a cluster population
and derive the LF for different cluster formation histories,
disruption mechanism and CIMFs. We will show that even if the initial
mass function of clusters is truncated, the brightest clusters in the
sample may will still be determined by sampling statistics as found by
\citet{2003dhst.symp..153W} and L02.

The paper is organized as follows: In \S~\ref{sec:data} we introduce
the data used for this study. In \S~\ref{sec:agemass} we present the
masses and luminosities as a function of log(age/yr) for three galaxies
in the sample. The LFs of cluster populations in five different
galaxies are presented in \S~\ref{sec:lf}. We introduce in
\S~\ref{sec:model} a cluster population model with which we can
reproduce LFs based on various variables. The relation between maximum
cluster mass and maximum cluster luminosity is discussed in
\S~\ref{sec:maxlum}. A discussion in presented in
\S~\ref{sec:discussion}. The conclusions are presented in
\S~\ref{sec:conclusions}.

\section{Data}
\label{sec:data}

\subsection{SMC and LMC}
The ages and masses of clusters in the LMC and SMC were derived by H03
from ground based broad-band photometry. The data was taken by the
Michigan Curtis Schmidt telescope by \citet{2002ApJS..141...81M} and
covers a large field of the LMC and SMC (11.0 kpc$^2$ and 8.3 kpc$^2$,
respectively).  Aperture photometry was used to derive $U,B,V$ and $R$
magnitudes. The colors and magnitudes were corrected for foreground
reddening ($E(B-V) = 0.13$ mag for the LMC and $E(B-V) = 0.09$ mag for
the SMC). The ages and masses where derived by comparing the colors
with the \citet{1999ApJS..123....3L} models. For more details on the
data reduction and age derivation we refer to H03.

\subsection{NGC~5236 and NGC~6946}
Data for these two galaxies was taken from L02 and we refer to that
  paper for details about the data reduction and analysis. Briefly,
  clusters were identified on $V$-band equivalent (F555W and F606W)
  {\it HST/WFPC2} images and aperture photometry was carried out using
  a 3 pixels aperture. Cluster candidates were confirmed by a visual
  inspection of the images. For NGC~5236 and NGC~6946 there were 1 and
  3 {\it HST/WFPC2} pointings available,
  corresponding to 7.1 kpc$^2$/53 kpc$^2$, respectively.
 
\subsection{M51}
We used the age, mass and luminosity data of more than 1000 clusters
in M51 found by \citet{2005A&A...431..905B}. Briefly summarized,
aperture photometry was performed on F336W, F439W, F555W, F675W,
F814W, F110W and F160W (roughly $U,B,V,R,I,J$ and $H$) images of the
{\it HST/WFPC2} and {\it HST/NICMOS3} cameras. The magnitudes and
colors were compared with the {\it GALEV} simple stellar population
models (\citealt{2003A&A...401.1063A}; \citealt{2002A&A...392....1S})
for different combinations of age and extinction. For detailed tests
of the accuracy of the ages and the dependence on photometric errors
and choice of models we refer to \citet{2005A&A...431..905B}. With 2
{\it HST/WFPC2} pointings we cover 69 kpc$^2$ of the inner region of
the galaxy.

\section{Masses and luminosities at different ages}
\label{sec:agemass}

\subsection{Maximum masses at different ages}

\subsubsection{Predictions}
The most direct way to study a possible truncation of the cluster mass
function at the high mass end is to look at the masses of clusters as
a function of the logarithm of the ages. When the cluster initial mass
function (CIMF) is a power-law distribution ($N\dr M \propto
M^{-\alpha^{\prime}}\dr M$) and not truncated, the maximum mass in the
sample is purely determined by the size-of-sample effect. A single age
distribution of clusters with a minimum mass $\mmin$ consisting of $N$
clusters, will statistically have a maximum mass of
$\mmax~=~N^{1/(\alpha^{\prime}-1)}\,\mmin$. H03 showed that the
maximum mass per logarithmic age bin should increase with log(age/yr)
(assuming a constant cluster formation rate), since each logarithmic
age bin corresponds to a longer time interval at older ages and hence
more clusters. The number of clusters at higher log(age/yr) increases
when assuming a constant cluster formation rate as: $N\,\dr \log(t)
\propto t\,\dr \log(t)$. So, the maximum mass of a sample of clusters
formed with a constant cluster formation rate should increase with age
($t$) as

\begin{equation}
\mmax\,\dr \log(t) \propto t^{1/(\alpha^{\prime}-1)}\,\dr \log(t).
\label{eq:mmax}
\end{equation}

\begin{figure}[!h]
    \includegraphics[width=8.cm]{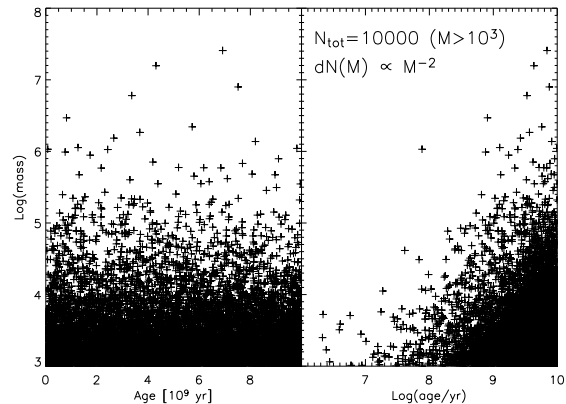} 

    \caption{Result of a Monte Carlo simulation where 10\,000 clusters
    are sampled from a power-law mass distribution and evenly
    distributed between $0 < t < 10^{10}$. In the {\bf left-hand} panel the
    masses are plotted {\it vs.} age and in the {\bf right-hand} panel {\it vs.}
    log(age/yr). In the latter bi-logarithmic plot an increase is
    visible with a slope of 1 (dashed line), which follows from Eq.~\ref{eq:mmax}.}
    
    \label{fig:linlog}
\end{figure}

This is illustrated in Fig.~\ref{fig:linlog}. We statistically generated  10\,000
clusters from a CIMF with exponent $-2$ between $6 < $ log(age/yr) $ < 10$. In
the left panel we plotted the results as a function of age and in the
right-hand panel as a function of log(age/yr). The increase of the
maximum mass predicted by Eq.~ \ref{eq:mmax} is only visible in the
right-hand panel.

\subsubsection{Observations}

\begin{figure*}[!t]
\begin{center}
    \includegraphics[width=18cm]{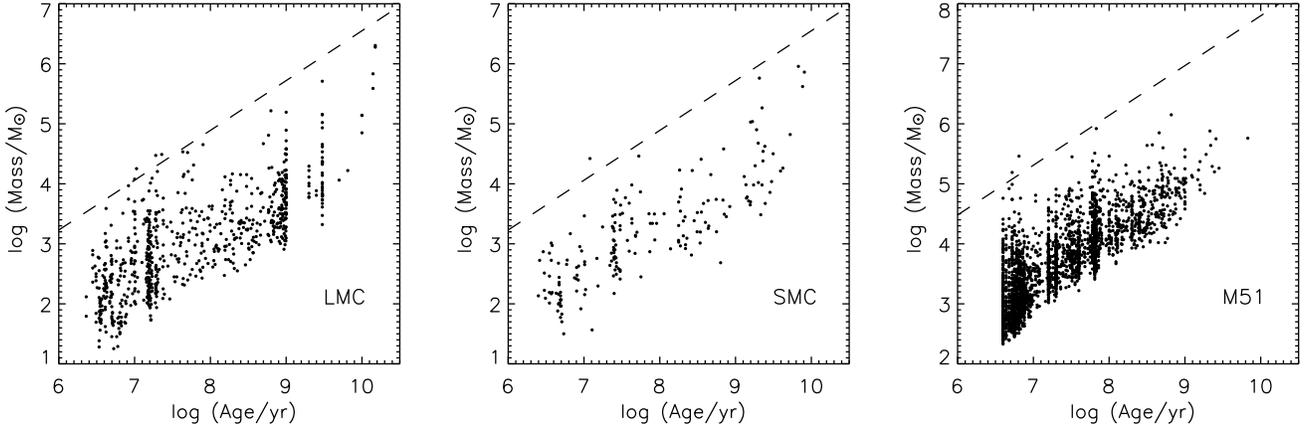}

    \caption{Cluster masses for different log(age/yr) for the LMC
    ({\bf left}), the SMC ({\bf middle}) and M51 ({\bf right}). All
    samples are limited by a detection limit. Since clusters fade when
    they age, the lower mass limit goes up in time. 
    the expected increase of the maximum cluster mass as a function of
    time based on the size-of-sample effect and an exponents for the
    initial cluster mass function of $\alpha^{\prime}= -2$.}

    \label{fig:age-mass}

\end{center}
\end{figure*}

For three galaxies in our sample (SMC, LMC and M51) we have ages
and masses of the individual clusters available from earlier
studies. For details on the derivation of the ages we refer to
\S~\ref{sec:data} and references therein. In Fig.~\ref{fig:age-mass}
we plot the masses of the clusters as a function of log(age/yr) for
the LMC (left), SMC (middle) and M51 (right). The completeness limit
increases with age to higher masses due to fading of clusters caused
by stellar evolution.  The first impression may be that the
unequal number of clusters in each age bin is incompatible with a
constant CFR. However, the observed age distribution is a function not
only of the cluster formation rate, but also of the detection limit and
disruption of clusters. In addition, the age-fitting method yields some
we irregularities. For examples all galaxies seem to have an over-density
of clusters at log(age/yr)=7.2. This is a known fitting artifact
(e.g. \citealt{2005A&A...441..949G}). Therefore, the unequal number of
clusters does not contradict a constant cluster formation rate.

The dashed line shows the slope of the predicted increase of the
maximum mass, based on $\alpha^{\prime} = 2$ and Eq.~\ref{eq:mmax}.  For
the LMC and SMC the maximum cluster mass as a function of log(age/yr)
can be well approximated by the size-of-sample prediction, as
was shown already by H03. 
For M51 however, we see that for ages higher than $10^8$ year, there
are more old high mass clusters predicted based on the size-of-sample
effect than observed. However, the observations show that there are no
clusters more massive than $\sim 10^6
\msun$.

\begin{figure}[!t]
\begin{center}
    \includegraphics[width=7.5cm]{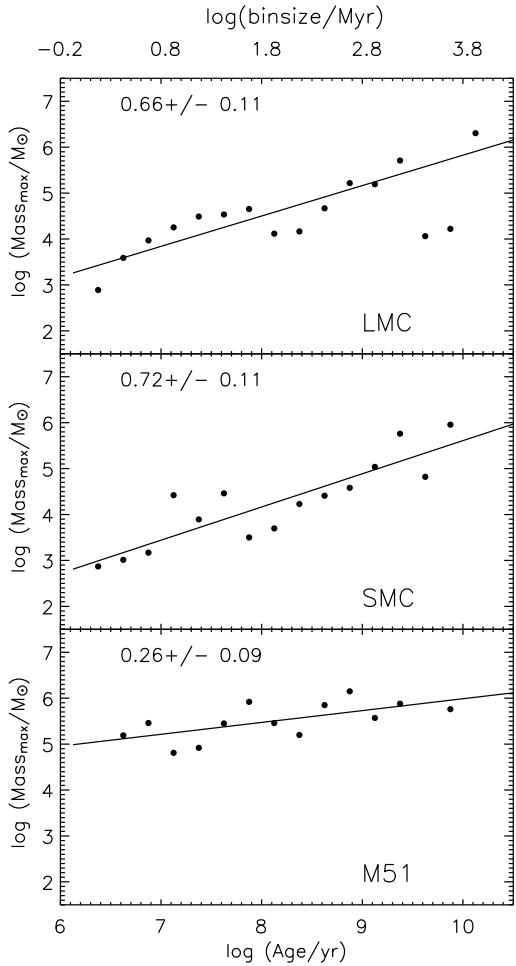}

    \caption{The most massive cluster (dots) in log(age/yr) bins of
    0.25 dex for all three galaxies. The fit to the most massive
    clusters is shown with the full line in all panels. The values of
    the slopes are indicated.  The top panel shows how the binsize
    increases with increasing log(age/yr), introducing the
    size-of-sample effect.}

    \label{fig:fitmm}
\end{center}    
\end{figure}

We also performed a linear fit to the maximum mass {\it vs.}
log(age/yr). This is shown in Fig.~\ref{fig:fitmm}. The most massive
cluster in every log(age/yr) bin is indicated with a dot. The fit is
shown with a full line.  The LMC clusters older than log(age/yr)=
9.5 were not included in the fit, because there seems to be an age gap
in the cluster distribution between 3 and 10 Gyr. For the SMC and LMC
the fit is consistent with the size of sample effect and a
CIMF exponent of $-\alpha^{\prime} = ~-2.4$, which was found already by
H03 for the SMC and LMC. This is a rather steep slope for the CIMF and
H03 showed that a fit to the first bins (log(age/yr) $<$ 8) yields a
value of $-\alpha^{\prime} = -2$. It might be that for higher ages we
are seeing a lack of old massive clusters. Note also that we here
assume a constant formation rate of cluster and no mass loss of the
clusters after formation. We will look at the effect of this
assumption in \S~\ref{subsubsec:massloss}. For M51 we find a much
shallower slope (0.26) than expected on account of the size-of-sample
effect. If this slope was caused by the mass function, it should have
an exponent of $-\alpha^{\prime} = -4.3$, which would be a much
steeper slope than found in any other study. In fact, the CIMF has
been determined for M51 by \citet{2003A&A...397..473B} from young
clusters and was found to be $-2.1\pm0.3$.

\subsubsection{The effect of mass loss, infant mortality rate and variable cluster formation rates}
\label{subsubsec:massloss}
In the conversion of the $\mmax$ {\it vs.} log(age/yr) relation to a
CIMF slope we assume that the clusters have not lost any mass after
their formation  and that the clusters were formed with a constant
formation rate. In reality, only 10-30\% of the clusters will
survive the initial 10 Myr, clusters will lose mass due to stellar
evolution (SEV) and tidal effects and the cluster formation rate (CFR)
might not be constant at all. To quantify these effect, we generated
a series of maximum masses consistent with the size-of-sample effect
and a mass function exponent of $-2$, assuming a constant CFR. We
then evolve each cluster mass as a function of its age according to
the analytical formulas introduced by \citet{2005A&A...441..117L}. The
mass of the cluster as a function of time can be well approximated by

\begin{equation}
M_p(t) = \left[(M_i\mu_{\rm sev})^\gamma - \gamma\,10^{4\gamma}\,\left(\frac{t}{t_4}\right)\right]^{1/\gamma},
\label{eq:mpresent}
\end{equation}
where $M_p(t)$ is the present mass of the cluster as a function of its
age, $M_i$ is the initial mass of the cluster, $\mu_{\rm sev} \equiv
M(t)/M_i$ is the fraction of remaining mass after mass loss due to stellar
evolution has been subtracted. The value of $\mu_{\rm sev}$ is based
on the {\it GALEV} models and analytically parameterized (see \citet{2005A&A...441..117L} for details). $t_4$ is the disruption time of
a $10^4
\msun$ cluster and $\gamma$ is a dimensionless index indicating the
dependence of the disruption time of clusters on the initial mass;
$t_{\rm dis}~\propto M_i^{\gamma}$. The value for $\gamma$ is found to
be 0.62 based on observations and $N$-body simulations
\citep{2005A&A...429..173L}.  Equation~\ref{eq:mpresent} describes the
mass of a cluster as a function of time based on a variable final
disruption time, taking mass loss due to stellar evolution and tidal
evaporation into account.

\begin{figure}[!h]
    \includegraphics[width=8.5cm]{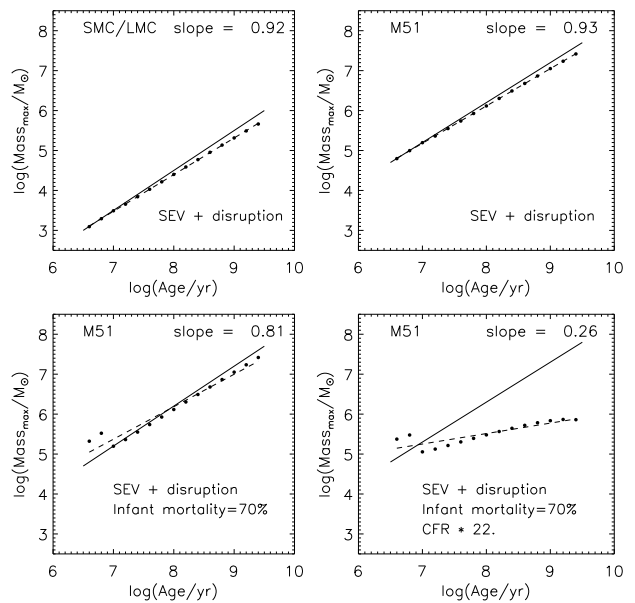} 

    \caption{A simulated set of initial maximum masses as a function
    of log(age/yr) (full line) for the SMC/LMC case and the M51 case
    with different effects.  The full line represent what is expected
    based on the size of sample effect and $-\alpha^{\prime} = -2$ and
    therefore has a slope of 1 (see Eq.~\ref{eq:mmax}). The initial
    masses were evolved in time taking into account mass loss due to
    stellar evolution and tidal effects. The evolved masses (dots) are
    fit and the slope is indicated with dashed lines.}
    
    \label{fig:maxexp}
\end{figure}

 We consider two scenarios: 1.) the SMC/LMC case and 2.) the M51
case. This is because the masses and disruption times are very
different. For the SMC/LMC case we choose the maximum mass to be
$\mmax = 10^3 \msun$ at log(age/yr) = 6.5, corresponding to the
observed maximum mass at young ages. For the M51 case we choose a
higher maximum mass; $\mmax = 5\times 10^4 \msun$ at log(age/yr) =
6.5, also based on the observed masses at young ages. We then assume a
typical disruption time of clusters for the SMC and LMC of $t_4 =
3\times 10^9$ yr and $t_4 = 2\times10^8$ yr for M51. The disruption
times are based on the observationally determined values of
\citet{2003MNRAS.338..717B} (SMC) and  \citet{2005A&A...441..949G}
(M51) and the predicted values of \citet{2003MNRAS.340..227B}
(LMC). In Fig.~\ref{fig:maxexp} we show the slope of the cluster
maximum masses as a function of log(age/yr) as predicted by the
size-of-sample effect with a full line and the shallower increase
after the clusters have been evolved in time for different
scenarios.  In the top panels we show the effect of stellar
evolution and disruption for the two different scenarios. The
disruption time of clusters in M51 is almost a factor of 15 shorter
than for clusters in the SMC/LMC, but since the masses are higher and
disruption is mass dependent ($t_{\rm dis}
\propto M_i^{0.62}$), the effect of the evolution is similar in both
cases.  Since this simple model can easily be extended to more
realistic scenarios, we now include the effect of a high infant
mortality rate in M51 in the bottom left panel of
Fig.~\ref{fig:maxexp}. \citet{2005A&A...431..905B} estimate that 70\%
of the clusters does not survive the first 10 Myr, independent of
their mass, so we increase the maximum mass for bins below 10 Myr with
a factor of 3.3 to simulate the higher masses at young ages. The
result is shown in the bottom left panel of Fig.~\ref{fig:maxexp}. The
inclusion of infant mortality makes the slope again slightly
shallower. However, still not even close to the observed value. The
last effect we consider, is a steady increase in the CFR. Since the
maximum mass is proportional to the CFR, an increase in the CFR could
result in more young massive clusters. To get to the slope of 0.26 as
observed in M51 (bottom panel of Fig.~\ref{fig:fitmm}), we apply an
additional increase in the CFR. We assume
that increase exponentially in time as CFR~$\propto t^\delta$. To get
to a slope of 0.26 as observed, we need $\delta = -0.5$, which means
the CFR has increased a factor of 22 in the last few Gyr. M51 is in
interaction with NGC~5195 which might have enhanced the CFR. In
earlier work, however, we have determined the CFR increase to be in
the order of a factor of 3-5 \citep{2005A&A...441..949G}, which is in
good agreement with independent determinations of the increase of the
star formation rate for these kind of interacting systems, based on
H$\alpha$ measurements \citep{2003A&A...405...31B}. Also, the increase
in the CFR seems to be more in bursts rather than an exponential
increase in time. A burst in the CFR will not affect the slope of
Fig.~\ref{fig:fitmm}, it will just show up as a peak. Worth noting is
that the two peaks of M51 (Fig.~\ref{fig:fitmm}) at log(age/yr)=7.8
and log(age/yr)=8.8. They coincide with the predicted moments of
encounter with NGC~5195, $\sim$75 Myr and $\sim$450 Myr ago \citep{2000MNRAS.319..377S}.

With this we show that part of the deviation from a slope of 1 in
Fig.~\ref{fig:fitmm} is caused by stellar evolution, cluster mass loss
due to tidal effects and infant mortality rate, but even the
short disruption time  and high infant mortality rate of M51 can
not explain the observed shallow increase of the maximum cluster mass
with log(age/yr) (Fig.~\ref{fig:fitmm}, bottom), without assuming an
unrealistically high increase in the CFR.

\subsection{The cluster mass function}

\begin{figure}[!t]
\begin{center}
    \includegraphics[width=8cm]{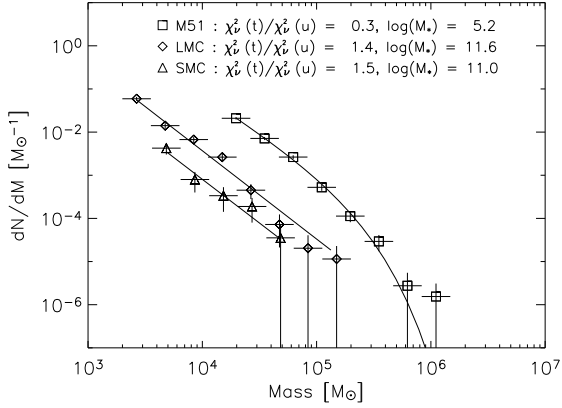}
    \caption{Mass functions of the LMC, SMC and M51 for an age
    and mass limited sample. The result of a truncated power-law fit
    is over-plotted and the corresponding $M_*$ is indicated. The
    $\redchisq$ of the truncated and untruncated power-law fits are
    compared  by showing the ratio.}
    \label{fig:mf}
\end{center}   
\end{figure}

If the CIMF is truncated, the total MF of clusters should also
show a truncation. As a check, we make the mass function, i.e. $\dr
N/\dr M$ as a function of mass, for the SMC, LMC and M51. Since the
detection limit cuts the sample at different masses
(Fig.~\ref{fig:age-mass}), we need to generate a mass and age limited
sample. We choose minimum mass and maximum age combinations of
$\left[\log(\mmin/\msun),\log({\rm age}_{\rm max}/{\rm yr}\right] =
[3.1,9], [3.5, 9]$ and $[4.2,8.7]$ for the LMC, the SMC and M51
respectively. The resulting mass functions are shown in
Fig.~\ref{fig:mf}. The three mass functions where fitted with a single
power-law, with variable slope and vertical scaling and with a power-law with
exponential cut-off: $N\dr M \propto
M^{-\alpha^{\prime}}\exp(-M/M_*)\,\dr M$, where the cut-off mass $M_*$
is an extra variable. This function is similar to the
\citet{1976ApJ...203..297S} function, which will later be used to fit
the luminosity function of clusters.  The $\redchisq$ results of the
truncated and untruncated power-law fits are compared by showing the
ratio. The fit of the truncated power-law is over-plotted. For the SMC
and LMC, however, this fit is similar to an untruncated power-law
fit. The M51 mass function is better fitted with a truncated
power-law. The values we find for $-\alpha^{\prime}$ are
$-2.05\pm0.01,-2.03\pm0.02$ and $-1.70\pm0.08$ for the SMC, LMC and
M51 respectively. The  value for M51 can be explained by 
mass dependent cluster disruption. If star clusters are formed with a
CIMF with index $-2$, then the total mass function will have the same
index, as long as the disruption time scale is longer than the maximum
age in the sample. For the SMC and LMC the maximum age in the sample
is $10^9$ yr, and the disruption time scale is $>> 10^9$ yr
\citep{2003MNRAS.340..227B,2003MNRAS.338..717B}. For M51, however, the
disruption time is shorter than the maximum age in the sample
\citep{2005A&A...441..949G}. Since the low mass clusters disrupt
faster than the high mass clusters, the mass function gets shallower
at older ages. Therefore we find that the MF has an index which is
greater than $-2.1$, which is the index of the CIMF of M51
\citep{2003A&A...397..473B}.

\subsection{Maximum luminosities at different ages}

\begin{figure*}[!t]
\begin{center}
    \includegraphics[width=18cm]{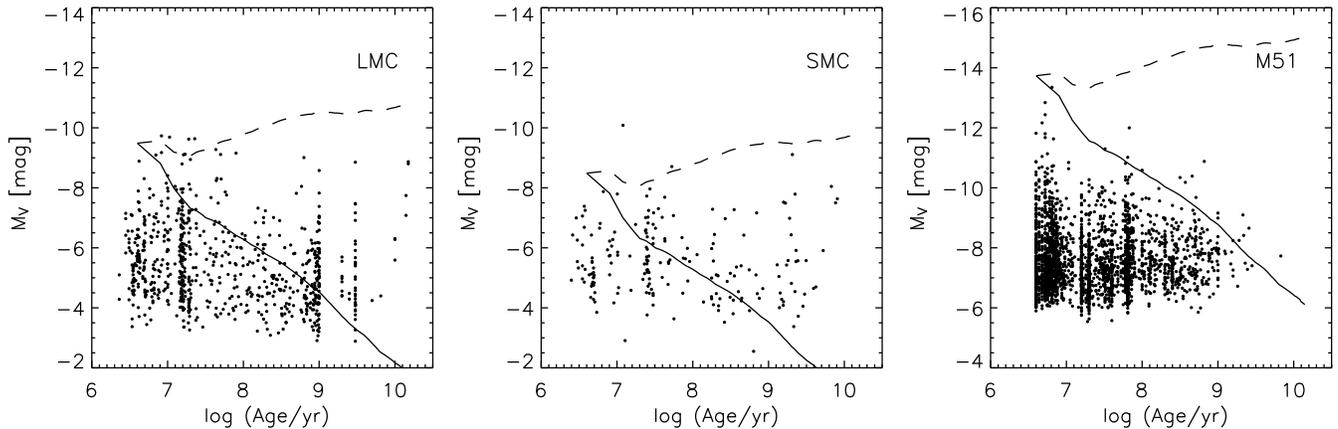} 

    \caption{Cluster absolute magnitudes ($M_V$) for different ages
    for the LMC ({\bf left}), the SMC ({\bf middle}) and M51 ({\bf
    right}). All samples are limited by a detection limit, which is
    why the lower parts of the diagrams are empty.  The solid line
    shows the fading of a cluster in the $V$-band based on the {\it
    GALEV} models.  The dashed line shows the expected luminosity of
    the maximum cluster mass as a function of time based on the
    size-of-sample effect and an exponents of the initial
    cluster mass function $\alpha^\prime = 2$.
    }  

    \label{fig:age-mv}

\end{center}
\end{figure*}

Another way of looking at the size-of-sample effect, is by plotting
the $M_V$ magnitudes of all clusters as a function of log(age/yr). In
Fig.~\ref{fig:age-mv} we plot the magnitudes for the three cluster
populations shown in Fig.~\ref{fig:age-mass}. The advantage of looking
at magnitudes instead of masses, although they are directly linked by
the age and mass dependent magnitudes from SSP models, is that this
figure helps to understand the LF, which will be discussed in
\S~\ref{sec:lf}. The solid line indicates the fading of a single mass
cluster based on the {\it GALEV} models, i.e. without tidal
disruption. We shifted the starting magnitude to the brightest
clusters at log(age/yr) = 6.5. The dashed lines now represent the
luminosity of the most massive cluster as a function of log(age/yr)
based on the size-of-sample effect and $\alpha^{\prime} = 2.0$. From the {\it GALEV}
simple stellar population we took the evolution of the magnitudes in
time for a single mass cluster with some reference mass ($M_{\rm
ref}$), which  is the maximum mass in the age range log(age/yr)
$<$ 7. We then add extra mass to $M_{\rm ref}$ as a function of
log(age/yr), following Eq.~\ref{eq:mmax}. The magnitude then depends
on the age and the exponent  of the CIMF as
\begin{equation}
M_V(t) \propto M_V(M_{\rm ref}, t)-2.5\,\log(t^{1/(\alpha^\prime-1)})
\label{eq:mvsos}
\end{equation}
where $M_{\rm ref}$ is a reference mass based on the observed maximum
mass at young ages, $M_V(t)$ is the maximum magnitude as a function of
age, $t$ is the cluster age and $\alpha^\prime$ is the exponent of the
CIMF. The dashed lines represent the maximum luminosity based on the
size-of-sample effect and Eq.~\ref{eq:mvsos}, for $\alpha^\prime = 2$.
The lower luminosity increases with log(age/yr) for M51. This
is since the clusters of M51 had to be detected in the F439W band in order
to make the observed sample. Since clusters fade more rapidly in the
filters bluer than the $V$ band, we miss some clusters at older ages
and faint magnitudes. Since we are here only interested in the maximum
luminosity, this will not affect our results.  When the colors
are used for cluster identification, these color differences have to
be taken in to account. 

The maximum luminosities of clusters in the LMC and SMC follow
the dashed lines better than the fading line of the single mass
cluster. The most luminous clusters in M51 seem to follow the fading
of the single mass $M_{\rm ref}$ better then the dashed lines which are based on
the size-of-sample prediction, which suggests that the underlying
maximum mass is truncated at that mass. 

Interestingly, \citet{1999ApJ...527L..81Z} made a similar plot for
the luminosities of clusters in the ``Antennae'' galaxies (their
Fig.~2). They also over-plot the luminosity evolution of single mass
clusters in time. Their maximum luminosities seem to follow the models
very well, which suggests that the maximum luminosity and hence the
maximum mass of clusters in the ``Antennae'' galaxies does not follow
the predicted size-of-sample relation of H03. {\it All this suggests
that the maximum mass of clusters in M51 and the
``Antennae''galaxies are not determined by size-of-sample effects, but
instead has a physical upper limit.}  The most luminous cluster in
these galaxies however, in general being a young clusters due to the
rapid fading in time of star clusters, {\it is} determined by
size-of-sample effects. This agrees with what was found by
\citet{2003dhst.symp..153W} and L02, who showed that the maximum
cluster luminosity in a galaxy depends on the total number of clusters
in the galaxy above a certain luminosity. We will continue with this
subject in \S~\ref{sec:maxlum}.

\section{The luminosity functions of five galaxies}
\label{sec:lf}

In this section we will present the LFs of (young) clusters in the
SMC, LMC, NGC~5236, NGC~6946 and M51. For each galaxy the number
of clusters as a function of $M_V$ was determined and a fit of the
form  $N\,\dr M_V~=~A\,10^{\beta\,M_V}\,\dr M_V$ was performed. The index
$\beta$ of the magnitude distribution relates to the exponent of the LF
as

\begin{equation}
-\alpha = -2.5\,\beta-1
\label{eq:ab}
\end{equation}
where $\alpha$ is the exponent of the LF ($N\,\dr
L~\propto~L^{-\alpha}\dr L$). In the following subsection we will
first present the LFs and fit a power-law function to all cluster
samples. In \S~\ref{subsec:fit} we will fit alternative functions and
compare them with the power-law fits to determine the best fit model
to the LFs. All results are summarized in Table~1.

\subsection{The observed luminosity functions}
\label{subsec:obs_lfs}

\subsubsection{SMC and LMC}
\label{subsec:lfsmclmc}
The LFs of the SMC and LMC are based on the data of H03. The LFs are
presented in the top two panels of Fig.~\ref{fig:lfs}. H03 report
lower limit magnitudes of $M_V = -3.5$ and $-4.5$ for the LMC and SMC,
respectively. Since the sample starts to be incomplete a bit
bright-ward of the cut-off magnitude, we used very conservative
completeness limits which are 1.5 mag brighter ($M_V = -5$ for the LMC
and $M_V = -6$ for the SMC). H03 only corrected for foreground
reddening ($E(B-V) = 0.09$ and 0.13 for the SMC and LMC respectively)
and we have no extinction estimates for the individual clusters
available. The detection limit is shifted the same amount as the
LF when correcting for foreground extinction.

\subsubsection{NGC~5236}
\label{subsec:lf5236}
The cluster sample of NGC~5236 is presented in L02. From the
completeness tests presented in that paper and a distance modulus of
27.9, we found a 90\% completeness limit of $M_V = -5.5$. We corrected
the measured magnitudes for galactic foreground extinction based on
\citet{1998ApJ...500..525S}. We used the sample which had spurious
sources removed after visual inspection.  A foreground extinction
correction of 0.218 mag was applied \citep{1998ApJ...500..525S}. The
sharp cut-off at high luminosities is because some of the bright
sources were saturated.

\subsubsection{NGC~6946}
The LF of clusters in NGC~6946 was presented in L02. We here use the
sample of clusters where spurious sources have been removed after
visual inspection. The 90\% completeness limit, following L02, is $V$
= 22.0 and a distance modulus of 28.9 mag was used. The magnitudes
were corrected for 1.133 mag foreground extinction based on
\citet{1998ApJ...500..525S}. The final LF can be seen in
Fig.~\ref{fig:lfs} (second from bottom).  The LF contains more
bright clusters than the previous ones. In addition the slope is
slightly steeper. The fact that LFs with brighter clusters are better
fitted with steeper power-laws was already noted by L02. In
\S~\ref{subsec:fit} we will compare this fit with a double power-law fit
and a Schechter function.

\subsubsection{M51}
\label{subsec:lfm51}
\begin{figure}[!h]
    \includegraphics[width=8.cm]{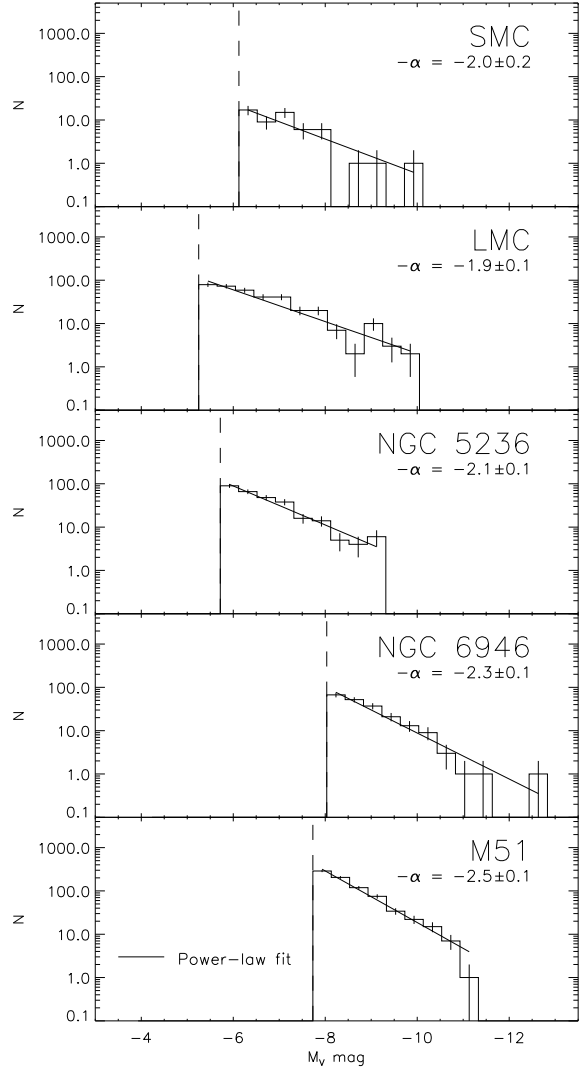}

    \caption{The (foreground extinction corrected) LFs for the five
    galaxies in our sample. In all graphs the dashed line represent
    the 90\% completeness limit. The solid lines are fits to the data
    above the 90\% completeness limit over the full range of the form
    $N\dr M_V = A\,10^{\beta\,M_V}\dr M_V$. From this the slope of the LF can
    be found using Eq.~\ref{eq:ab}. }
    \label{fig:lfs}
\end{figure}

For the LF of M51 the data from \citet{2005A&A...431..905B} is
used. We refer to this paper for detailed explanation on data
reduction and source detection. Ideally, we select clusters for the LF
independent of the applied age fitting routine. The best criterion for
this would be to select on extended objects brighter than the 90\%
completeness limit in F555W. Since we only found reliable radius
estimates for $\nrextended$ objects, we selected the observed clusters
that passed our fitting criteria from \citet{2005A&A...431..905B},
e.g. the sources that were well fit by a cluster model. Since one of
our source selection criteria is that the object has to be above the
90\% completeness limit in at least four bands, it is of importance
which band we use. Comparing the SEDs for different ages of the {\it
GALEV} models to the 90\% completeness limits in all filters, we found
that all clusters are always above our 90\% completeness limit in
F555W, F675W and F814W once they are detected in the F439W filter. If
we apply the F555W 90\% cut to the sample, our four-filter criterion
will cut the sample sometimes because the source has to be visible in
F439W as well, depending on the age. To be sure that for all ages our
sample is limited by a cut in the F555W band only, we add the maximum
color difference F439W -- F555W from the {\it GALEV} models for
clusters between $4\times10^6$ and $10^9$ years to our F555W 90\%
completeness limit. This makes the cut in this filter 0.6 mag brighter
than the 90\% completeness limit of 22.7 mag. A distance modulus of
M51 for 29.62 mag was used \citep{1997ApJ...479..231F} and a
foreground extinction correction of 0.117 mag was applied
\citep{1998ApJ...500..525S}. The LF for all clusters brighter than
22.0 mag in F555W is shown in the bottom panel of
Fig.~\ref{fig:lfs}.  As was the case for the LF of NGC~6946, we
also fit a power-law which is steeper than the power-law CIMF index
$-2$ of \citet{1999ApJ...527L..81Z}. We will consider different
functions in
\S~\ref{subsec:fit}.

\subsection{Fitting other functions to the luminosity functions}
\label{subsec:fit}

The brighter LFs of NGC~6946 and M51 are fitted with power-law
functions with steeper slopes ($-\alpha \simeq-2.4$) than the fainter ones
($-\alpha \simeq-2$). When a single power-law function is fitted to faint
sides of the LFs of NGC~6946 and M51, we find similar values as for
the SMC, LMC and NGC~5236. This suggests that the LFs turn over at
the bright side, something what was found for the ``Antennae''
galaxies by W99 as well. In order to quantify the deviation from a
single power-law, we fitted {\it all} LFs of \S~\ref{subsec:obs_lfs}
with a double power-law function with variable break point and a
Schechter function. The Schechter function is generally used for the
LF of galaxies. It has a power-law nature on the faint end and a
exponential drop at the bright side. In terms of magnitudes the
Schechter function can be expressed as

\begin{eqnarray}
N(M_\lambda){\rm d}M_\lambda & = &  0.4\,\ln(10)\,N_*\,10^{0.4\,(-\alpha+1)\,(M_*-M_\lambda)}\times\nonumber\\\nopagebreak
& & \exp\left[-10^{0.4\,(M_*-M_\lambda)}\right]\,\,{\rm d}M_\lambda ,
\label{eq:schechter}
\end{eqnarray}
where $M_\lambda$ is the magnitude in a band with central wavelength
$\lambda$, $N_*$ is a normalization number, $\alpha$ is the exponent
of the power-law part of the LF and $M_*$ is the characteristic point of
the function. $M_*$ is the equivalent of $L_*$ when using the
Schechter function to fit the LF of galaxies.

\begin{figure}[!h]
    \includegraphics[width=8.cm]{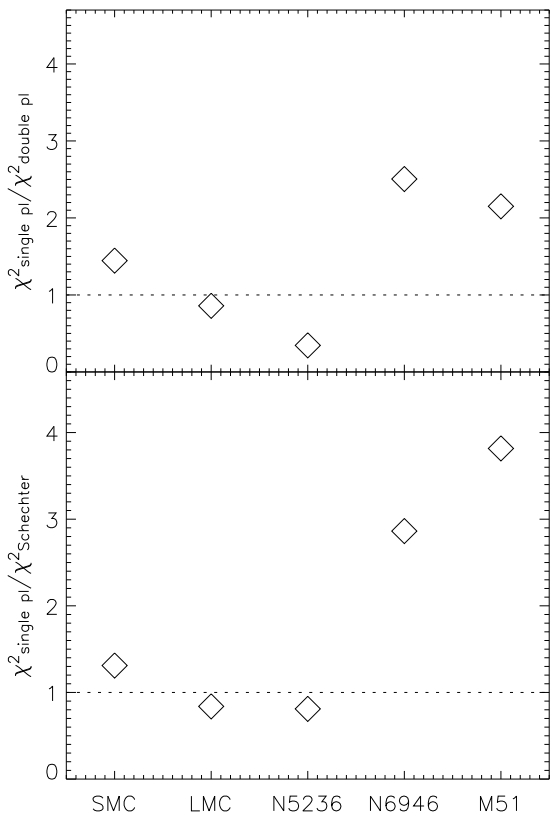}

    \caption{Comparison between the $\redchisq$ results of the
    different functions that were fitted to all LFs. {\bf Top:} Ratio
    of $\chi^2_{\nu,\,{\rm single \,power-law}}$ over
    $\chi^2_{\nu,\,{\rm double\,power-law}}$ for the five galaxies of
    which we have LFs. {\bf Bottom:} Ratio of $\chi^2_{\nu,\,{\rm
    single \,power-law}}$ over $\chi^2_{\nu,\,{\rm Schechter}}$. The
    LFs of NGC~6946 and M51 are better fitted by a double power-law
    distribution or a Schechter function, as compared to a single
    power-law distribution.}

    \label{fig:chisq}
\end{figure}

To compare the different function fits for all five galaxies, we
compare the reduced $\chi^2$ values of the different fits. In the top
panel of Fig.~\ref{fig:chisq} we show the ratio of $\chi^2_{\nu,\,{\rm
single \,power-law}}/\chi^2_{\nu,\,{\rm double\,power-law}}$ for the
five LFs. In the bottom panel we show the ratio of $\chi^2_{\nu,\,{\rm
single \,power-law}}/\chi^2_{\nu,\,{\rm Schechter}}$. The $\redchisq$
comparison shows that the LFs of NGC~6946 and M51 are better fitted
with a function that drops at the bright end,  although from the
values in Table~1 and the $\redchisq$ comparison mentioned, a single
power-law distribution can not be excluded with a high significance level.

\begin{figure}[!h]
    \includegraphics[width=8.cm]{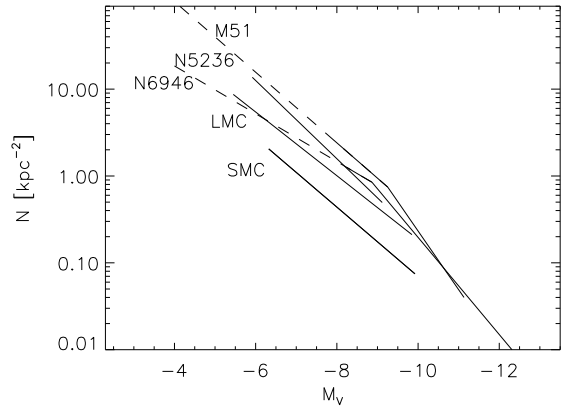}

    \caption{The best fit result to all five LFs. The number of
    clusters is normalized to the observed area of the galaxies in
    kpc$^2$. The full line is the region where data points were
    available from observations. The dashed lines are
    extrapolations. The two galaxies which are better fitted by a
    double power-law (NGC~6946 and M51), are on top and the only
    galaxies with cluster brighter than $M_V = -10$ mag.}

    \label{fig:fits}
\end{figure}

In Fig.~\ref{fig:fits} we show the best fits to the LFs for all five
galaxies, based on their $\redchisq$ results. The number of clusters
is normalized to the observed area for fair comparison.  We
extrapolated the faint side of the LFs of NGC~6946 and M51 with a
dashed line, this to show that these galaxies on the faint side have
similar slopes as the other three cluster LFs. Another interesting
fact is that NGC~6946 and M51 are on top, due to a high
CFR. Therefore, they are also the only galaxies with cluster brighter
than $M_V = -10$ mag.

\subsubsection{A closer look at NGC~6946 and M51}
\label{subsec:fitbend}

\begin{figure*}[!t]
\begin{center}
    \includegraphics[width=15cm]{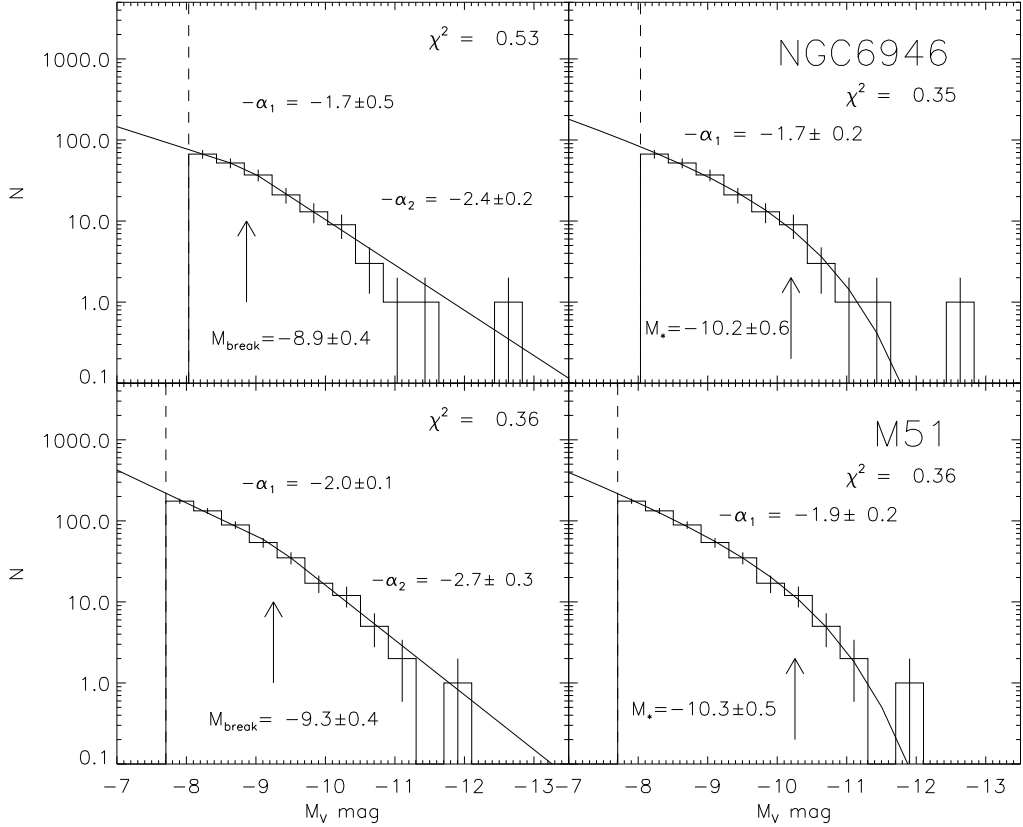} 

    \caption{The extinction corrected LFs for NGC~6964 ({\bf top}) and
    M51 ({\bf bottom}). Results for a double power-law fit ({\bf left})
    and a Schechter function ({\bf right}) to the same data are shown.}

    \label{fig:4panel}
\end{center}
\end{figure*}

\begin{figure}[!h]
    \includegraphics[width=8.cm]{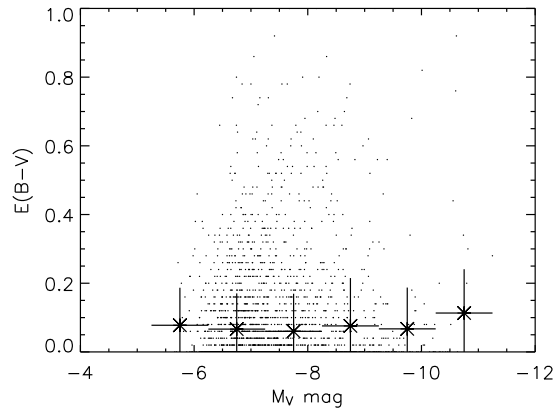}

    \caption{$E(B-V)$ {\it vs.} luminosity for all clusters of M51,
    taken from \citet{2005A&A...431..905B}. The average values at different magnitudes
    are indicated.}

    \label{fig:ebv}
\end{figure}
\subsubsection*{NGC~6946}

 To get a realistic idea of the location of the breakpoint, which we
can compare with other galaxies, we need to take into account possible
reddening. L02 did not determine reddening for the individual objects,
but only mentions the Galactic foreground reddening. For the direction
of NGC~6946 this is $A_V = 1.13$ and we corrected the $V$ magnitudes
of each cluster for this amount. The breakpoint was determined by
fitting double power-laws with a variable breakpoint, exponents and
normalization constant. We found the break to be at $M_V =
\breakngcs$, with an exponent faint-ward of the break of $-\alpha_1 =
\slopengcsone$ and $-\alpha_2 =
\slopengcstwo$ bright-ward of the break. The (foreground) extinction
corrected LF with the double power-law fit can be seen in the top left
panel of Fig.~\ref{fig:4panel}. W99 observe similar behavior for the
LF of the ``Antennae'' galaxies. When they fit two power-laws, they
find the breakpoint around $M_V = -10.3$, depending on the different
samples they selected.

We also fitted a Schechter function \citep{1976ApJ...203..297S},
which has a more gradual bend and it is usually used to fit the LF of
galaxies. This was also done for the clusters in the ``Antennae''
galaxies by W99. 

The top right panel of Fig.~\ref{fig:4panel} shows the result of the
Schechter function fit to the (foreground) extinction corrected LF of
NGC~6946. The bend in the Schechter function occurs at $M_* =
\mstarngcs$, which is  1.2 mag fainter than the $M_*$ of the
``Antennae'' galaxies ($M_* = \mstarant$). The break point in the
double power-law distribution occurs 1.2 mag fainter as compared to
the ``Antennae''.

\subsubsection*{M51}
For M51, we corrected all magnitudes for the extinction values we
acquired from the age fitting (see \citet{2005A&A...431..905B} for
details). This is important, since extinction effects might also
affect the shape of the LFs. \citet{2005A&A...431..905B} found that
young clusters have on average higher extinction values than the older
ones. L02 looked at the effect of differential extinction on the
LF. He found that it makes the LF steeper, but only when there is
already a cut-off in the mass distribution on the high mass end. If
the mass function is not truncated, differential extinction does not
affect the shape of the LF, since then still power-laws with the same
index would be added, resulting in a power-law with that index. In
case there is a relation between the luminosity and extinction, for
example because we can see more intrinsically bright clusters, the
bright part of the LF will be extinct ed more and this could result in
a drop. To check this, we plotted the available extinction and
luminosities from \citet{2005A&A...431..905B} in
Fig.~\ref{fig:ebv}. We averaged the extinction for different
luminosity bins, and there is no relation between the two
parameters. So the fits to the LF of M51 will have a higher
reliability than the ones of NGC~6946, since there individual source
extinction is not known.

When we fit a double power-law distribution, the location of the bend
is at $M_V = \breakmf$. The exponents on the faint and bright side of
the breakpoint are $-\alpha_1 =
\slopemfone$ and $-\alpha_2 = \slopemftwo$ respectively (bottom left
panel of Fig.~\ref{fig:4panel}). These values are very similar to
those found for  NGC~6946. The bottom right panel
of Fig.~\ref{fig:4panel} shows the result of a Schechter function fit
to the M51 clusters. The exponent is $-\alpha = -1.9$
and the bend is at $M_*~=~\mstarmf$.

The results of all fits to the LFs are summarized in Table~1. We added
the values found by W99 for the ``Antennae'' galaxies as a
comparison. The two exponents of the double power-law fits are very
similar between NGC~6946, M51 and the ``Antennae''. The breakpoint in
the case of the double power-law fit, however, occurs about 1.2 mag
brighter in the ``Antennae'' galaxies. The value for $M_*$ is also
about 1 mag brighter compared to NGC~6946 and M51.

In the next section we will attempt to explain the shape of the LFs,
in particular the bend observed for NGC~6946, M51 and the ``Antennae''
galaxies.

\begin{table*}[!t]
\centerline{Table 1. Properties of the observed LFs.}
\vskip 1mm
\begin{center}
\begin{tabular}{lllllll}
\hline\hline

Galaxy \hspace{1cm}  &$-\alpha_1$              &$-\alpha_2$                &Break             &$M_*$                 & Foreground extinction ($A_V$)      \\  
                     &                         &                           &      (mag)       &      (mag)           &      (mag)      \\  \hline
SMC                  &$\slopesmc$              &-                          &-                 &-                     & 0.09     \\
LMC                  &$\slopelmc$              &-                          &-                 &-                     & 0.13     \\
NGC~5236             &$\slopengcf$             &-                          &-                 &-                     &0.218      \\
NGC~6946             &$\slopengcsone$          &$\slopengcstwo$            &$\breakngcs$      &$\mstarngcs$          &1.133   \\
M51$^1$              &$\slopemfone$            &$\slopemftwo$              &$\breakmf$        &$\mstarmf$            &- \\ 
{\it``Antennae''$^2$}&{\it $\slopeantone$}     &{\it $\slopeanttwo$}       &{\it $\breakant$}  &{\it $\mstarant$}    &-          \\ \hline

\multicolumn{5}{l}{$^1$ Corrected for individual cluster extinction.}\\
\multicolumn{5}{l}{$^2$ Values from W99 for their sample of extended sources.}\\

\end{tabular}
\end{center}
\end{table*}

\section{A cluster population model}
\label{sec:model}

Our goal is to understand why we observe a bend in the luminosity
function of NGC~6946, M51 and the ``Antennae'' galaxies and not in the
other galaxies in the sample.  The luminosity function (LF) of a star
cluster population is a combination of multi-age cluster initial mass
functions, which all have age dependent mass to light ratios. Clusters
will lose mass in time due to stellar evolution and tidal effects,
which will also affect the luminosity as a function of age. When there
are no ages and masses of the individual clusters available, an
alternative way to look at how the LF is built up is to model the LF
based on various physical input parameters. In this section we will
analytically generate the LF of a star cluster population.

\subsection{Creating a synthetic cluster population}
\label{subsec:createpop}

\subsubsection{Set-up}
The synthetic cluster population is based on an analytical model
presented in \citet{2005A&A...441..949G}. In that work it was used
to predict ages and masses of the cluster population of M51. The
LF follows directly from these predictions since the
combination of age and mass can be converted to luminosity using
simple stellar population models. 

The model creates a series of
cells equally spaced in log($t$) and log($M$). Then a CIMF and
formation rate 
are chosen.  Based on these choices each cell
is assigned a weight ($w(t,M_i)$). To acquire a CIMF
with exponent $-\alpha^{\prime}$ and a constant formation rate, $w(t,M_i)$ can be
written as
\begin{equation}
 w(t, M_i) = (t/t_{\rm min})\times(\alpha^{\prime} -1 ) \times (M_i/M_{\rm max})^{1-\alpha^{\prime}}
\end{equation}
where $w(t, M_i)$ is the weight assigned to a cell with age $t$ and
 mass $M_i$, $M_{\rm max}$ is the mass of the most massive cell in the
 simulation, $t_{\rm min}$ is the age of the youngest cell in the
 simulation and $\alpha^{\prime}$ is the exponent of the mass
 function. The weight is now a function of age and initial mass, where
 the most massive cell in the youngest bin has a weight of one, and
 corresponds to 1 cluster per log($t$)--log($M$) bin. This value is
 arbitrary and can change since later we will scale the total
 population to a different number. When $-\alpha^{\prime} = -2$ is
 chosen, the weight depends on age and mass simply as: $w(t,M_i)
 \propto t/M_i$. When the cells are then binned in age, mass or
 luminosity, the weights of each cell are counted and the bin values
 now represent number of clusters. The main advantage of using these
 weight assigned cells spread equally in log($t$) and log($M$), is
 that it would otherwise be very time consuming to generate cluster
 populations with a well sampled mass function; the amount of clusters
 needed for that is too high. Also, the number of points per bin is
 now constant and it is straight-forward to create a variety of
 populations with different formation rates, disruption time scales
 etc. in a short time. The model from \citet{2005A&A...441..949G}
 is extended  in this work  to create LFs. To this end the age and mass
 dependent magnitudes of clusters are taken from the {\it GALEV} SSP
 models based on a Salpeter IMF. When generating a LF, the total
 weight of the cluster population is scaled to a realistic number of
 clusters. Then only bins with weights greater or equal to
 one are kept. In this way the predicted maximum luminosity follows
 directly from the model, depending on the number of clusters (=total
 weight) and the input variables.

\subsubsection{Include stellar evolution and cluster disruption}
\label{subsubsec:sevdis}

In a recent study \citep{2005A&A...441..117L} it was shown that
there is a simple analytical description of the evolution of the mass
of a single cluster as a function of time (see Eq.~\ref{eq:mpresent},
\S~\ref{subsubsec:massloss}). It takes into account the effect of mass
loss due to stellar evolution, based on the mass loss predicted by the
{\it GALEV} models and it includes mass loss due to tidal evaporation
based on the $N$-body results of \citet{2003MNRAS.340..227B}.  The
mass as a function of time according to these analytical formula
agrees very well with the predictions from $N$-body
simulations \citep{2005A&A...441..117L}. We assume stars of all type
are lost due to disruption. In reality more low mass stars will leave
the cluster. This might slightly affect the magnitudes as a function
of time.

\subsection{The luminosity function of a multi-age population}
\label{subsec:multiage}

In the first step to understand the LF we generate an
analytic cluster population with log($t$) ranging from 7 to
9. Cells are spread equally in log($t$) and log($M)$ space
and weights are assigned to correct for the logarithmic binning in
age and to create a mass function with  $-\alpha^{\prime} = -2.0$. Cluster
masses were assigned between 10$^3 \msun$ and $10^{12}
\msun$. The upper mass was chosen unrealistic high, representing the
lack of a physical upper limit to the most massive cluster in the mass
function. The total weight is then scaled to 1000 clusters. In
principle this step is not necessary, it tells us what the most
luminous cluster in the sample is, since we will only use bins with values greater
than one. (Note that due to the weights we use and the final scaling
there are cells with weights much smaller than one.) Every cell is
assigned a magnitude based on the {\it GALEV} models. The exponent of
the resulting LF for this multi-age population is always $-\alpha = -2.0$.
Although clusters with log($t$) = 9 have faded considerably, the final
LF is still an addition of power-laws with exponents of $-2$, with no limit
to the maximum mass and hence the maximum luminosity. For this
reason, changes in the formation rate as well as bursts will have no
effect on the slope. Mass dependent disruption will slightly change
the slope (\S~\ref{subsubsec:disruption}).

\subsection{The luminosity function without a maximum mass}
\label{subsec:notruncation}

When the CIMF is not truncated, the mass of the most massive cluster
observed will increase with increasing log(age/yr) due to the
size-of-sample effect (\S~\ref{sec:agemass}; H03). The most massive
cluster will depend on age as $\log(M_{\rm max}) \propto \log(t)$,
when $-\alpha^\prime = -2$. The dotted line in Fig.~\ref{fig:bend}
takes this increase of the mass {\it and} the fading of the most
massive cluster as a function of log(age/yr) into account. In two dex
in age the maximum mass will also increase with two dex if the slope
of the mass function is $-2$. This means that the most massive cluster
at $10^9$ years will be a factor of 100 more massive and hence 5 mag
brighter than the one at $10^7$ years. Since clusters fade about 4 mag
between $10^7$ and $10^9$ years, these two effects almost cancel
out. Not only the mass of the most massive cell will increase, but
also the second most massive etc. Therefore, the LFs at different ages
will all have clusters sampled between the detection limit and more or
less the same maximum luminosity. So, the slope of the integrated
luminosity function will be close to the slope of the CIMF for almost
all luminosities, which we found already in
\S~\ref{subsec:multiage}. This effect can be seen in
Fig.~\ref{fig:age-mv} for the LMC and the SMC clusters. At all
ages we find clusters between the detection limit and more or less the
same maximum luminosity. Integration over all ages will yield a LF
which has about the same slope as the mass function. This has always
been assumed to be logic, but here we show that this is only
the case when there is no maximum mass.

\subsection{Truncation of the mass function}
\label{subsec:truncation}

\begin{figure}[!t]
    \includegraphics[width=8.cm]{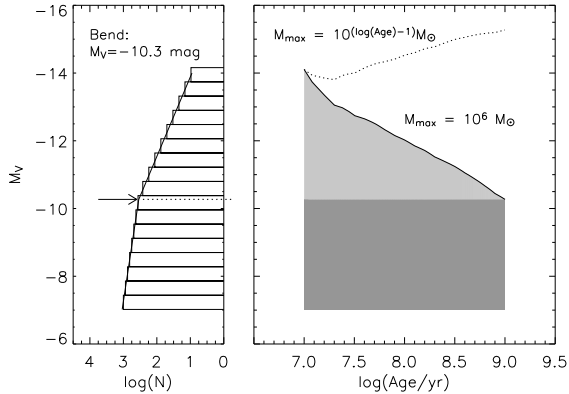}

    \caption{{\bf Right}: Schematic picture of the two regions (dark
    and light grey) contributing to the LF when the underlying CIMF is
    truncated. The dark shaded region contributes to the faint end of
    the LF and the light shaded region to the bright end. The line
    indicates the brightness of a $10^6 \msun$ cluster. The dotted
    line indicates the brightness of the most massive cluster taking
    size of sampling effects into account. {\bf Left}: Resulting
    LF. }

    \label{fig:bend}
\end{figure}

L02 has already shown that when generating a multi-age population with
a truncation at the high mass side, the LF gets steeper. This can be
understood as follows: when the underlying mass function is truncated
at some upper mass, the brightest cluster will be the most massive
cluster in the youngest age bin.  This holds when the MF is
sampled up to the truncation limit in the youngest age bin. The older
clusters have faded due to stellar evolution. Hence at the brightest
end of the LF only the youngest clusters will contribute. Going to
fainter magnitudes, more and more ages will contribute to the LF. At
the luminosity of the most massive cluster in the oldest age bin and
fainter, all ages will contribute, and therefore the slope faint-ward
of that brightness will be the slope of the CIMF, similar to what we
found in
\S~\ref{subsec:multiage}. This is illustrated in
Fig.~\ref{fig:bend}. The decreasing line in the right hand panel
represents the luminosity of a 10$^6 \msun$ cluster between
$10^7$ and $10^9$ years. The dark shaded region is where all clusters
of all ages contribute to the same part of the LF. The light shaded
region represent the clusters contributing to the bright end of the
LF.

\begin{figure}[!h]
    \includegraphics[width=8.cm]{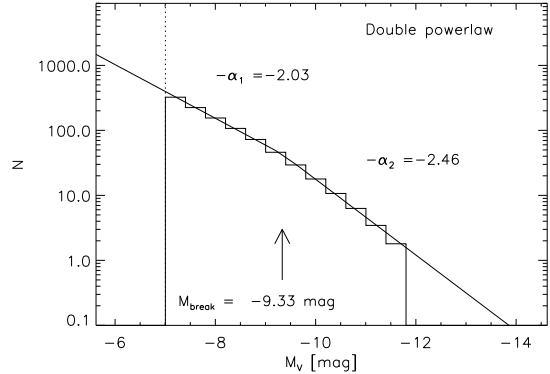}\\
    \includegraphics[width=8.cm]{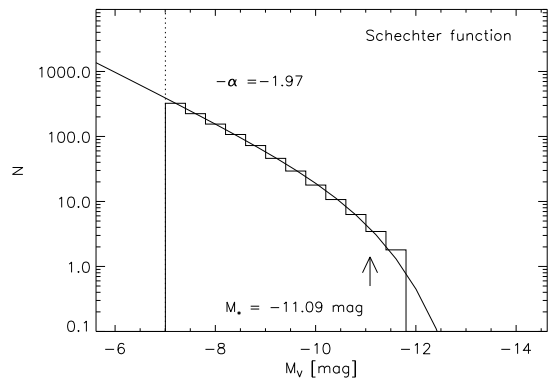}

    \caption{LF of a synthetic cluster population with a CIMF with
    exponent -2, a mass function truncation at $\mmax = 10^6 \msun$
    and a detection limit at $M_V = -7$. The top panel shows a double
    power-law fit, where the slope bright-ward of the bend is steeper
    than the underlying mass function. The bottom panel shows a fit
    with a Schechter function.}

    \label{fig:v_mod_bend}
\end{figure}

When there is a physical mechanism that does not allow clusters more
massive than $M_{\rm max}$ to be formed, there should be an observable
bend at $M_\lambda = M_\lambda(M_{\rm max},t_{\rm max})$. With our
model  we now generate a cluster population
with ages between $10^7$ and $3\times10^9$ yr and masses between
$10^2$ and $10^6$ and apply a detection limit at $M_V = -7$. The age
range is based on the observed age range in M51. We excluded the
clusters younger than log(age/yr)=7, since a large fraction of these
clusters will not survive the first 10 Myr
(e.g. \citealt{2003ARA&A..41...57L}; \citealt{2005A&A...431..905B}). We apply no disruption or extinction. The total
weight of this simulation is scaled to 1000 clusters. 
The brightest cluster now is the first bin at the bright side with a
weight larger than 1. The resulting LF is shown in
Fig.~\ref{fig:v_mod_bend}. The faint part has an exponent of $-\alpha
= \slopemodone$, which is expected, since the underlying mass function
has an exponent of $-\alpha^{\prime} = -2$. On the bright side the
slope is steeper ($\slopemodtwo$) and the break between the two
power-laws is at $M_V = \breakmod $ mag, which is close to the
magnitude of a $10^6 \msun$ cluster with an age of $3\times10^9$ yr
($M_V(t=3\times10^9$ yr$;M=10^6 \msun) = -9.1$ mag) . For NGC~6946 and
M51 the observed break occurs at $M_V = \breakngcs$ mag and $M_V =
\breakmf$ mag respectively implies a maximum cluster mass of
$6.9\times10^5
\msun$ and $1.0\times10^6 \msun$. For comparison, the bend in the LF of the ``Antennae''
galaxies at $M_V = -10.3$ corresponds to a maximum mass of
$\sim2.5\times10^6 \msun$, assuming the oldest cluster in the sample is
also 3 Gyr. We also fit a Schechter function (Eq.~\ref{eq:schechter})
to the model as was done for the clusters in NGC~6946, M51 and the
``Antennae'' galaxies. When we fit this
function to our model with $\alpha, N_*$ and $M_*$  as variables we get
a very good fit (see bottom panel of
Fig.~\ref{fig:v_mod_bend}). As was the case with the Schechter fit to
the observed LF of NGC~6946 and M51, the bend now occurs brighter than when
breaking the LF into two power-law distributions. When comparing the
bend points from the Schechter functions to the model, we find maximum
cluster masses of $4.4\times 10^5 \msun$, $4.8\times 10^5 \msun$ and
$1.3\times 10^6 \msun$ for NGC~6946, M51 and the ``Antennae''
respectively. We note that the errors in the fitted break points and
Schechter bends are large ($\sim$ 0.5 mag), which corresponds to a factor
1.5 in mass. It is therefore hard to place a hard limit on the
upper mass of the CIMF.



\subsection{Impact of different initial conditions}
\label{subsec:initial}

Our assumed initial conditions of \S~\ref{subsec:truncation}
might affect our results and in particularly the location of a break
in the LF. We therefore created several cluster populations, with
different initial conditions.

\subsubsection{Varying cluster formation rate}
\label{subsubsec:cfr}
For simplicity we have taken the cluster formation rate (CFR) to be
constant over 3 Gyr in \S~\ref{subsec:truncation}. Especially when
comparing with M51 and the ``Antennae'' galaxies this might be an over
simplification. We generated models with bursts in the CFR, as is
probably more realistic for M51 \citep{2005A&A...441..949G}. Bursts
with increased CFRs of factors of 2-3 do not affect the location of
the bend. For comparison with the LF of the ``Antennae'', we increased
the CFR the last 160 Myr, based on the results of
\citet{2003ApJ...599.1049W}. For a CFR which was 3/5/10 times higher
the last 160 Myr as before, the bend occurs 0.3/1.5/1.8 mag
brighter. This would make our observed maximum masses 1.3/4/5 times
lower.  As was shown in \S~\cite{subsubsec:massloss}, the
assumption of a constant CFR will always make it hard to directly
compare the model with the data.

\subsubsection{The stellar IMF}
\label{subsubsec:imf}
Our SSP models assume a Salpeter IMF. When a Kroupa was assumed, the
corresponding mass derived from a bend in the modeled LF will be about
a factor of 2 lower. To explain the bend location in M51, NGC~6946 and
the ``Antennae'' galaxies, our derived maximum masses will also be a
factor of 2 lower.

\subsubsection{The age range}
\label{subsubsec:agerange}
We assumed clusters in the age range of $10^7 <$ age $< 3\times 10^9$
years, based on the observed cluster ages in M51. When we would take
$10^9$ years as the maximum age range, the location of the bend would
be 0.8 mag brighter. The maximum masses of our observations would be a
factor of 2 lower in that case. If we would change the youngest
cluster in our model from $10^7$ years to $4\times10^6$ years, the
modeled location of the bend does not change. However, when we assume
that only 20\% of the youngest clusters (e.g. age $< 10^7$ years)
survive the gas removal phase (e.g. \citealt{2004ASPC..322..399F};
\citealt{2005A&A...431..905B}), it would place the bend 0.4 mag
fainter. That would make our maximum masses derived from observations
a factor of 1.5 more massive. In addition, clusters will probably
loose 50-80\% of their stars in this phase
\citep{2002MNRAS.336.1188K}, which will make {\it all} our derived
{\it initial} maximum masses a factor 2-5 more massive.

\subsubsection{The effect of disruption}
\label{subsubsec:disruption}
Applying cluster disruption according to Eq.~\ref{eq:mpresent} affects
only the LF at the faint end. The low mass clusters are disrupted
faster and that will result in a shallower mass and hence LF at higher
ages. Since the young clusters also contribute at these brightnesses,
it is hard to quantify this affect. When applying the short cluster
disruption time of M51 ($t_4 = 10^8$ years), we find an exponent of
the LF on the faint side of $\alpha = 1.7$ and the location of the
bend does not change. It could explain why in the ``Antennae''
galaxies slopes significantly shallower than -2 are reported (W99)
after various steps to removes stellar contamination. Also this value
is close to the observed faint slope of the LF of NGC~6946 and it
could explain why the slopes of the LF of the SMC and LMC are
shallower than -2.

\subsection{Testing the bend scenario in different filters}

\begin{figure*}[!t]
    \includegraphics[width=18cm]{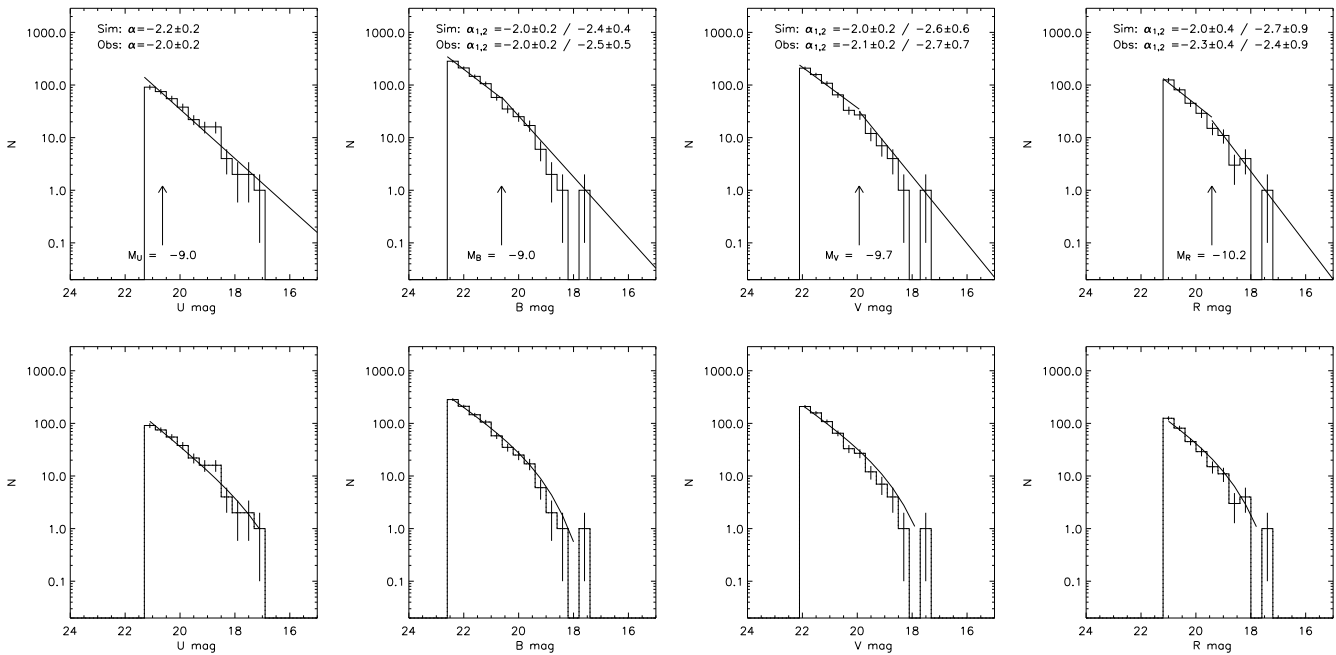} 
    
    \caption{The LF of 705 clusters with ages between
    $10^7$ and $10^9$ year in M51 in the F336W, F439W, F555W and F675W
    bands. Over plotted with solid lines are fits to modeled luminosity
    function, based on a clusters IMF with exponent -2.1, equal number of
    clusters and a truncation of the mass function at $\mmax =
    10^{5.9}$. {\bf Top:} Fits to the model are two power-laws, with
    the break at $M_\lambda = M_\lambda(M_{\rm max},t_{\rm max})$. The
    bend occurs to brighter magnitudes going to redder filters, since
    a cluster of age $10^9$ years is brighter in the redder filters. {\bf Bottom:} Same data,
    but now the fits to the model are Schechter functions.}

    \label{fig:lf-allbands}
\end{figure*}

If the location of the bend in the LF is related to the most massive
cluster in the oldest age bin (\S~\ref{subsec:truncation}), it should
appear at brighter luminosities in the red filters because old
clusters have red colors. With this knowledge, we can test our theory on
the cluster sample of M51, for which we have multiple filters
available. With our model we can now predict the LF and a possible
bend in all filters. We used a constant formation rate and a
disruption time of clusters of $t_4 = 10^8$ years, as was found
by \citet{2005A&A...441..949G}. The CIMF was truncated at the high
mass side. The clusters were generated and evolved according to method
described in
\S~\ref{subsec:createpop}. The bend will occur at different magnitudes
for the different filters because the magnitude of the maximum cluster
will be different in the different filters.  The result is shown in
Fig.~\ref{fig:lf-allbands}. The top and bottom panel show the data in
the F336W, F439W, F555W and F675W filters, roughly equal to $U, B, V$
and $R$. In the top panel we plot the data and over plotted double
power-law fits to the model. The break point between the two
power-laws was based on the magnitudes of $\mmax$ in these filters at
$10^9$ years according to the {\it GALEV} SSP models.  The predicted
bend is indicated in each band in the top panel. 

The break moves to brighter luminosities for redder filters as
predicted, this is because clusters of $3\times10^9$ year are in general
redder than younger clusters. For example, the $B-V$ color at $10^9$
year is 0.7 mag, so the bend in the $V$-band occurs 0.7 mag brighter
than in the $B$-band. In the observed LF also the bend seems to move
to brighter luminosities when going to redder filters. In the
$U$-band, the bend is predicted 0.8 mag brighter than the detection
limit. We thus only see the steep part of the LF. The observed LF of
the $U$-band seems also to be a single power-law, although a slight
bend is visible just bright-ward of the detection limit. The steep part
is much shallower in the F336W band than in the other filters due to
the rapid fading in that filter. The data and the model agree very
well on this. The bottom panel of Fig.~\ref{fig:lf-allbands} shows the
same data, but now the models were fitted with Schechter functions
(Eq.~\ref{eq:schechter}). The fit to the model ends at the bright side
at the last bin which had a value of 1 or higher. This means that this
is the luminosity of the brightest cluster in the simulation. The
brightest cluster in the model agrees well with the brightest cluster
in the data.

The bend in the ``Antennae'' occurs $\sim5$ mag brighter in the
Ks-band \citep{mengel05} as in the $V$-band (W99). The $V-K$ model
color of a cluster of 3 Gyr is around 3.2 mag. The $V$-band
observations of W99 are not corrected for extinction, making the bend
in $V$ fainter and most probably the observations by \citet{mengel05}
are hampered by blends, making the bend brighter in Ks.

\subsection{Effect of a bend in the mass function}

\begin{figure}[!t]
    \includegraphics[width=8cm]{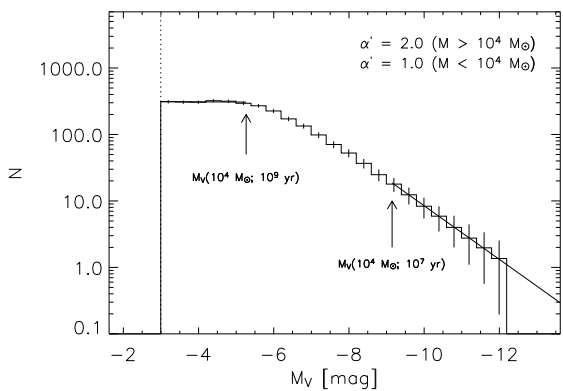} 
    
    \caption{Simulated LF of clusters which are formed with a constant
    formation rate between 10$^7$ and 10$^9$ years and a bend in the
    CIMF. Below $10^4 \msun$ the mass function has an exponent of
    $-\alpha^{\prime} = -1$ and above $10^4 \msun$ the exponent is
    $-\alpha^{\prime} = -2$. The mass function has no truncation at
    high masses. At the bright side the LF has the a slope of $\alpha = 2$ and on the very faint side the slope is $\alpha = 1$.}

    \label{fig:bendmf}
\end{figure}

The observed bend in the LF of the ``Antennae'' galaxies has been
interpreted as a bend in the mass function (W99;
\citealt{1999A&A...342L..25F}). With our model we can predict how a
bend in the CIMF would affect the LF. We generate a cluster sample
between $10^7$ and $10^9$ years with a constant formation rate. We put
a bend in the CIMF at $M = 10^4 \msun$, where the exponent is
$-\alpha^{\prime} = -1$ below the bend and $-\alpha^{\prime} = -2$ on
the high mass end. The resulting LF is shown in
Fig.~\ref{fig:bendmf}. On the very faint side the LF has the same
slope as the underlying MF below $10^4 \msun$. Note that a slope of
the LF of $-\alpha = -1$ shows up as a flat part in the magnitude
distribution due to the conversion between the slope of the magnitude
distribution and the slope of the LF (Eq.~\ref{eq:ab}). On the bright
side the LF has the same slope as the MF above $10^4 \msun$ ($-\alpha
= -2$). The transition region is caused by the fading of a $10^4
\msun$ cluster between $10^7$ and $10^9$ years. When we would fit two
power-laws over the whole LF, the break would be somewhere at $M_V =
-6$ mag, i.e. much to faint to observe a bend in the MF at $10^4
\msun$.

\subsection{Bend {\it vs.} no bend}
So if the bend in the LF (\S~\ref{subsec:truncation}) can be
explained with a truncated mass function, why do the other galaxies in
our sample (SMC, LMC, and NGC~5236,
\S~\ref{subsec:lfsmclmc}-\ref{subsec:lf5236}) not show this bend? The
answer is sampling statistics.  If the bend is a physical feature
occurring at a given mass, then it should always occur at a
predictable luminosity (assuming the age distribution is known). Of
the galaxies studied here, NGC~6946 and M51 (and the ``Antennae'') are
the only ones where the LF is sampled above $M_V=-10$. In the SMC, LMC
and NGC~5236 we do not sample the LF up to such bright magnitudes, so
if the truncation occurs at the same mass as in N6946/M51 we would not
be able to see it.

\section{The maximum cluster luminosity}
\label{sec:maxlum}

Now that we have shown that there are arguments to believe that the
CIMF is truncated, it is interesting to compare how this might affect
the maximum cluster luminosity in a
galaxy. \citet{2003dhst.symp..153W} plotted the maximum cluster
luminosity in a set of galaxies is a function of the number of
clusters above $M_V = -9$ mag.  A fit to the data is close to the
expected relation based on the size-of-sample effect and an exponent
of the luminosity function of $\alpha$ = 2. L02 has proposed a
relation between the maximum cluster luminosity and the product of the
observed area ($A$) and the area-normalized SFR ($\Sigma_{\rm
SFR}$) of the galaxy, assuming that the cluster formation rate is
proportional to the SFR. The increase of the maximum cluster
luminosity with the product $A*$SFR is consistent with the observed
increase.

\begin{figure}[h]
    \includegraphics[width=8cm]{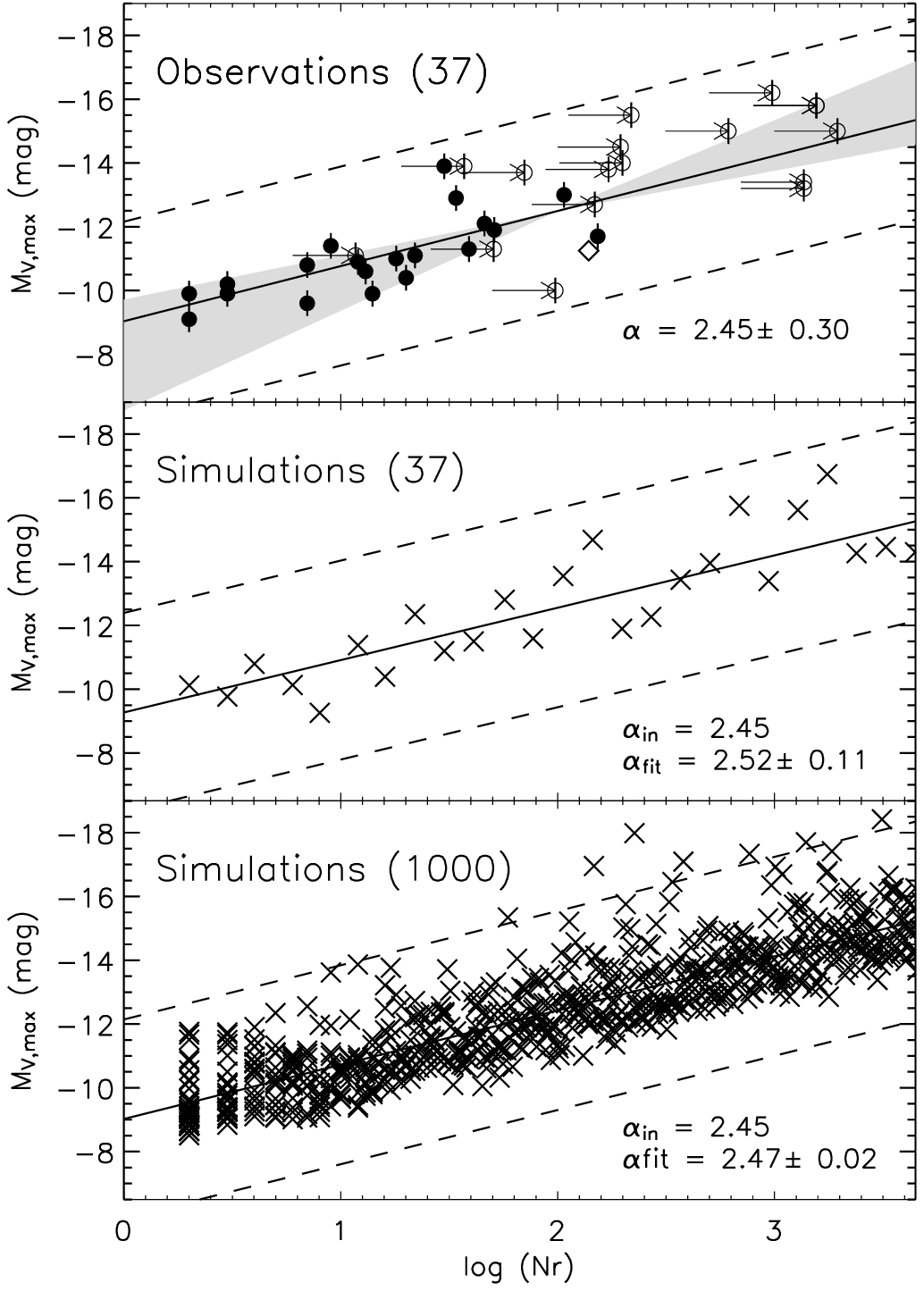}

    \caption{The maximum cluster luminosity in a galaxy {\it vs.} the
    number of clusters above a certain magnitude limit. {\bf Top:}
    Observed maximum luminosities for 37 galaxies, compiled from a
    sample by \citet{2003dhst.symp..153W} with open circles and
    \citet{2000A&A...354..836L} shown in full circles. The open
    circles are shifted 0.2 dex, since that sample is for clusters
    with $M_V < -9$ mag, while the sample of
    \citet{2000A&A...354..836L} is for clusters with $M_V < -8.5$. The
    grey area represents the expected range based on the
    size-of-sample effect and the observed range of $\alpha$ in this
    study ($-2.8 < -\alpha < -1.8$). The full line is a fit to the
    data. Some special cases are indicated as: M51 ($\diamond$),
    NGC~1569 ($\bigtriangleup$). The {\bf middle} and {\bf bottom}
    panel show simulated maximum cluster luminosities for similar
    number of galaxies as the observations and a large number
    respectively. The simulations were based on $-\alpha = -2.45$.  In
    all three plots the 3$\sigma$ deviation predicted by L02 is
    over-plotted with a dashed line.}
    
     \label{fig:maxlum}
\end{figure}

In the top panel of Fig.~\ref{fig:maxlum} we show the brightest
cluster as a function of log($N(M_V < -8.5$)) from the sample of
\citet{2000A&A...354..836L} with filled circles and
\citet{2003dhst.symp..153W} with open circles. The sample of
\citet{2003dhst.symp..153W} has been
shifted to the right with 0.2 dex, which is the expected number of
clusters missed between $M_V = -8.5$ and $M_V =-9.0$ based on a
luminosity function with exponent $\alpha = 2$. The $\Delta \log(N)$
in a certain magnitude interval relates to the slope of the LF as

\begin{equation}
\Delta \log[N(\Delta M)]~=~0.4\,(\alpha-1)\,\Delta M ,
\end{equation}
where $\Delta M$ is the magnitude interval and $\alpha$ is the slope
of the LF. We have updated the data for NGC~1569, based on the data of
\citet{2000AJ....120.2383H}. This was an outlying point in
\citet{2003dhst.symp..153W}, but we found more clusters in
\citet{2000AJ....120.2383H}.

We have over-plotted the expected increase in maximum cluster
luminosity with $\log(N)$ assuming the luminosity function is a single
power-law distribution and based on a range of exponent
($1.7~<~\alpha~<~2.7$), which is the observed range of slopes in this
work. The fit to the closed circles is shown with the solid line,
which falls nicely within the predicted area and corresponds to a
exponent of the LF of $\alpha = 2.45\pm0.30$. Note that although this
value is slightly higher than found in \S~\ref{subsec:fit}, it is
still within the 1$\sigma$ error. Also we here use an indirect method
to determine the slope of the LF, based on a small number of data
points.

L02 has quantified the expected scatter in the maximum cluster
luminosity. The 1$\sigma$ deviation is 1.04 mag from the mean and is
independent of the number. We over-plotted the $3\sigma$
deviations. To show the effect of the expected scatter, we show a
simulated set of maximum cluster luminosities as a function of
$\log(N)$ in the middle and bottom panels of
Fig.~\ref{fig:maxlum}. The input exponent of the simulated LF was the
value found from the fit to the observations (2.45). The same number
of simulated maximum cluster luminosities as was observed is shown in
the middle panel and a larger number in the bottom panel. The
simulated luminosities show that luminosities close to the $3\sigma$
error line are expected even with the low number of galaxies. The
3$\sigma$ scatter is a little bit higher on the bright side of the
fit, which is because of the power-law nature of the LF function.

For the galaxies studied here, we find that the luminosity of the
brightest cluster can be accounted for by sampling statistics. This
is in contrast with the most massive cluster. This can be understood
by looking at Fig.~\ref{fig:age-mass} and Fig.~\ref{fig:age-mv}. There
we can see that the most luminous cluster in each galaxy is always a
young cluster (log(age/yr) $\simeq$ 7). The most massive cluster,
however, is usually in the older bins, since these cover a larger
time in which it is more likely to form a more massive clusters.  The
physical upper limits we have derived in this work are not reached
when sampling a CIMF during 10$^7$ years.

\section{Discussion}
\label{sec:discussion}

A truncation of the cluster mass function raises the question what
physical mechanism could be responsible for this cut-off. An answer
might be present in the mass spectrum of giant molecular clouds
(GMCs). \citet{1997ApJ...476..166W} have shown that there is a
distinct cut-off at the high mass end of the mass function of GMCs in
the Milky Way. There are no clouds observed above $6\times10^6
\msun$, while there are $\sim$100 GMCs expected above that cut-off,
based on statistical arguments. This might be caused by shear effects
which do not allow clouds to grow larger than a certain size. The
size and the mass of clouds are directly related by the virial
theorem. Assuming that star clusters are formed with a star formation
efficiency of $\epsilon =0.01-0.1 $ and that one single GMC produces
one cluster, this implies it is not possible at this moment to form
clusters more massive than $0.6-6.0 \times10^5
\msun$ in the Milky Way. The Milky Way has formed a few massive
clusters in the last few Myr, like Westerlund 1 with a mass of $\sim
10^5 \msun$ \citep{2005A&A...434..949C} and the Arches and
Quintuplet clusters in the Galactic center with masses of around $10^4
\msun$ \citep{1999ApJ...525..750F}.

A rough estimate can show that we would have expected more
massive objects. Westerlund 1 is the most massive young cluster in the
Milky Way disk. Of course we have to look at the size-of-sample that
was used to discover it. Westerlund 1 is about 4.5 kpc away from the
sun \citep{2005A&A...434..949C}. Based on the most recent catalog of
open clusters by \citet{2005A&A...438.1163K},
\citet{2005A&A...441..117L} estimated the masses of clusters in the
solar neighborhood. Within 600 pc from the sun, the catalog is
complete until 200 $\msun$, where about 100 clusters are found above
that mass. The initial mass of the most massive cluster observed is
estimated to be $3\times10^4 \msun$. The size of sample predicts
$\mmax = N\,\mmin = 100*200 \msun = 2\times10^4 \msun$. So within the
solar neighborhood the mass of most massive cluster is determined by
the size-of-sample effect. Assuming that the number density of
clusters is constant until Westerlund 1, we can predict the expected
maximum mass, based on statistical grounds. Based on the distance of
4.5 kpc, the expected number of clusters above 200 $\msun$ is about
($4500^2/600^2 = 56$) times larger. This implies we would expect a
maximum mass of $\mmax = 1.5\times10^6
\msun$. Although Westerlund 1 is a very massive cluster with
$10^5\,\msun$, we would have expected at least one cluster above
$10^6\,\msun$. Given that Westerlund 1 is highly obscured, it is
unlikely that a 10 times more  massive object would not have been found yet
within a distance on 4.5 kpc from the sun. In fact, the mass of
Westerlund 1 might be a reasonable upper limit, given that the most
massive GMC is $6\times10^6\,\msun$ \citep{1997ApJ...476..166W} and a
reasonable star formation efficiency of a few percent.

It is not clear what causes the cut-off of the cloud mass
function. The maximum cluster mass seems to be environment dependent,
since in the ``Antennae'' galaxies clusters of a few times $10^6
\msun$ are being formed. This could be because the maximum GMC mass is
higher \citep{2003ApJ...599.1049W}, or because the star formation
efficiency is much higher. The star formation efficiency is expected
to be higher in high pressure environments
\citep{1997ApJ...480..235E}, which is the case in shock induced star
formation environments like galaxy mergers
\citep{2005sdlb.proc..143S}. The super massive cluster W3 in NGC~7252,
with a dynamical mass of $10^8
\msun$ \citep{2004A&A...416..467M}, exceeds the limits we
observe by almost two orders of magnitude. It was probably formed in
the merger process of two gas-rich spirals and seems to be the tip of
a continues power-law distribution \citep{1997AJ....114.2381M}.

 As a comparison, the ``Antennae'' galaxies have not merged their
nuclei yet. Simulations of the merging of two gas-rich spirals
\citep{1991ApJ...370L..65B} have shown that in the third encounter the
gas disks merge with relative velocities of more than 500
$\kms$. Perhaps more massive clusters can be formed in these merging
stages, due to the merging of clusters. Clusters also seem to
form in complexes of multiple clusters and stars, with similar
properties of their progenitor GMCs
\citep{2005astro.ph..8110B}. \citet{2005MNRAS.359..223F} have shown by
numerical simulations that these complexes can collapse and form a
single, diffuse, ultra-massive object, which might be the way W3 and
other very massive clusters can be formed.  During the merger
shear-free regions exist, like in the {\it overlap-region} in the
``Antennae''galaxies, where GMCs can grow bigger and more massive than
in the galaxies we have studied here.

\section{Conclusion}
\label{sec:conclusions}

We have compared observed star cluster luminosity function in five
galaxies with analytical cluster population models. Our main results
can be summarized as follows:

\begin{itemize}
\item  
there are no clusters in M51 more massive than $M \simeq 1\times10^6
\msun$, although they are predicted by the size-of-sample effect. 
When comparing the maximum cluster mass in increasing log(age/yr)
bins, the LMC and SMC cluster population show an increase consistent
with the size-of-sample of effect. The cluster population of M51,
however, shows a much shallower increase. This suggests a physical
upper limit to the masses of clusters M51, although the shallow
increase can also be reproduced by a combined effect of cluster
disruption, infant mortality and an increasing cluster formation rate.
\item 
 when comparing the $\redchisq$ results of different function fits
to the five galaxies in our sample, we find that the LF of the SMC,
the LMC and NGC~5236 can be well approximated by a power-law ($N\,\dr
L \propto L^{-\alpha}\,\dr L$), with $1.9 < \alpha < 2.1$, while the
LF of NGC~6946 and M51 are slightly better approximated with a double
power-law or Schechter function.
\item 
 when fitting a double power-law function to the LF of NGC~6946
we find a break point at $M_V$ = $\breakngcs$ mag. Faint-ward of the
bend a power-law with exponent $\slopengcsone$ can be
fitted. Bright-ward of the bend an exponent of $\slopengcstwo$ is
found. The LF can also be well fitted by fit a Schechter function with
a bend at $M_V$ = $\mstarmf$ mag.
\item 
the LF of M51 shows a  break
at $M_V$ = $\breakmf$ mag. Faint-ward of the bend a power-law with
exponent $\slopemfone$ can be fitted. Bright-ward of the bend an
exponent of $\slopemftwo$ is found. The LF of M51 is also well fitted
by a Schechter function with a bend at $M_V$ = $\mstarmf$ mag.
\item 
the cluster LFs can be reproduced with a synthetic cluster population
model. The bend in the LF of NGC~6946, M51 and ``Antennae'' galaxies
can be explained with a truncation of the cluster mass function at
$\mmax = 0.5-1\times10^6 \msun$ (M51/NGC~6946) and $1.3-2.5\times10^6 \msun$
(``Antennae'').

\end{itemize}

In a follow-up study \citep{gieles05b} we present an improved LF of
star clusters in M51 based on {\it HST/ACS} data, taken as part of the
Hubble Heritage project.

\begin{acknowledgements}
We are very grateful to Deidre Hunter who kindly provided the data of
the SMC and LMC clusters. We like to thank Carsten Weidner and Pavel
Kroupa for useful discussion. We acknowledge research support from and
hospitality at the International Space Science Institute (ISSI) in
Berne (Switzerland), as part of an International Team
programme. Finally, we thank an anonymous referee for very
constructive comments on the manuscript, which have improved the
paper.
\end{acknowledgements}

\bibliographystyle{aa}

\begin{thebibliography}{49}
\expandafter\ifx\csname natexlab\endcsname\relax\def\natexlab#1{#1}\fi

\bibitem[{{Anders} \& {Fritze-v.~Alvensleben}(2003)}]{2003A&A...401.1063A}
{Anders}, P. \& {Fritze-v.~Alvensleben}, U. 2003, \aap, 401, 1063

\bibitem[{{Barnes} \& {Hernquist}(1991)}]{1991ApJ...370L..65B}
{Barnes}, J.~E. \& {Hernquist}, L.~E. 1991, \apjl, 370, L65

\bibitem[{{Bastian} {et~al.}(2005{\natexlab{a}}){Bastian}, {Gieles}, {Efremov},
  \& {Lamers}}]{2005astro.ph..8110B}
{Bastian}, N., {Gieles}, M., {Efremov}, Y.~N., \& {Lamers}, H.~J.~G.~L.~M.
  2005{\natexlab{a}}, \aap, 443, 79

\bibitem[{{Bastian} {et~al.}(2005{\natexlab{b}}){Bastian}, {Gieles}, {Lamers},
  {Scheepmaker}, \& {de Grijs}}]{2005A&A...431..905B}
{Bastian}, N., {Gieles}, M., {Lamers}, H.~J.~G.~L.~M., {Scheepmaker}, R.~A., \&
  {de Grijs}, R. 2005{\natexlab{b}}, \aap, 431, 905

\bibitem[{{Baumgardt} \& {Makino}(2003)}]{2003MNRAS.340..227B}
{Baumgardt}, H. \& {Makino}, J. 2003, \mnras, 340, 227

\bibitem[{{Bergvall} {et~al.}(2003){Bergvall}, {Laurikainen}, \&
  {Aalto}}]{2003A&A...405...31B}
{Bergvall}, N., {Laurikainen}, E., \& {Aalto}, S. 2003, \aap, 405, 31

\bibitem[{{Bik} {et~al.}(2003){Bik}, {Lamers}, {Bastian}, {Panagia}, \&
  {Romaniello}}]{2003A&A...397..473B}
{Bik}, A., {Lamers}, H.~J.~G.~L.~M., {Bastian}, N., {Panagia}, N., \&
  {Romaniello}, M. 2003, \aap, 397, 473

\bibitem[{{Boutloukos} \& {Lamers}(2003)}]{2003MNRAS.338..717B}
{Boutloukos}, S.~G. \& {Lamers}, H.~J.~G.~L.~M. 2003, \mnras, 338, 717

\bibitem[{{Clark} {et~al.}(2005){Clark}, {Negueruela}, {Crowther}, \&
  {Goodwin}}]{2005A&A...434..949C}
{Clark}, J.~S., {Negueruela}, I., {Crowther}, P.~A., \& {Goodwin}, S.~P. 2005,
  \aap, 434, 949

\bibitem[{{de Grijs} {et~al.}(2003{\natexlab{a}}){de Grijs}, {Anders},
  {Bastian}, {Lynds}, {Lamers}, \& {O'Neil}}]{2003MNRAS.343.1285D}
{de Grijs}, R., {Anders}, P., {Bastian}, N., {et~al.} 2003{\natexlab{a}},
  \mnras, 343, 1285

\bibitem[{{de Grijs} {et~al.}(2003{\natexlab{b}}){de Grijs},
  {Fritze-v.~Alvensleben}, {Anders}, {Gallagher}, {Bastian}, {Taylor}, \&
  {Windhorst}}]{2003MNRAS.342..259D}
{de Grijs}, R., {Fritze-v.~Alvensleben}, U., {Anders}, P., {et~al.}
  2003{\natexlab{b}}, \mnras, 342, 259

\bibitem[{{Elmegreen} \& {Efremov}(1997)}]{1997ApJ...480..235E}
{Elmegreen}, B.~G. \& {Efremov}, Y.~N. 1997, \apj, 480, 235

\bibitem[{{Fall}(2004)}]{2004ASPC..322..399F}
{Fall}, S.~M. 2004, in ASP Conf. Ser. 322: The Formation and Evolution of
  Massive Young Star Clusters, 399--+

\bibitem[{{Feldmeier} {et~al.}(1997){Feldmeier}, {Ciardullo}, \&
  {Jacoby}}]{1997ApJ...479..231F}
{Feldmeier}, J.~J., {Ciardullo}, R., \& {Jacoby}, G.~H. 1997, \apj, 479, 231

\bibitem[{{Fellhauer} \& {Kroupa}(2005)}]{2005MNRAS.359..223F}
{Fellhauer}, M. \& {Kroupa}, P. 2005, \mnras, 359, 223

\bibitem[{{Figer} {et~al.}(1999){Figer}, {Kim}, {Morris}, {Serabyn}, {Rich}, \&
  {McLean}}]{1999ApJ...525..750F}
{Figer}, D.~F., {Kim}, S.~S., {Morris}, M., {et~al.} 1999, \apj, 525, 750

\bibitem[{{Fritze-v.~Alvensleben}(1999)}]{1999A&A...342L..25F}
{Fritze-v.~Alvensleben}, U. 1999, \aap, 342, L25

\bibitem[{{Gieles} {et~al.}(2005{\natexlab{a}}){Gieles}, {Bastian}, {Lamers}, \&
  {Mout}}]{2005A&A...441..949G}
{Gieles}, M., {Bastian}, N., {Lamers}, H.~J.~G.~L.~M., \& {Mout}, J.~N. 2005{\natexlab{}},
  \aap, 441, 949

\bibitem[{{Gieles} {et~al.}(2006{\natexlab{b}}){Gieles}, {Larsen}, {Scheepmaker},
  {Bastian}, {Haas} \& {lamers}}]{gieles05b}
{Gieles}, M., {Larsen}, S.~S., {Scheepmaker}, R.~A. {et~al.} 2006{\natexlab{}},
  accepted for \aap\ Letters (astro-ph/0512298)


\bibitem[{{Hunter} {et~al.}(2003){Hunter}, {Elmegreen}, {Dupuy}, \&
  {Mortonson}}]{2003AJ....126.1836H}
{Hunter}, D.~A., {Elmegreen}, B.~G., {Dupuy}, T.~J., \& {Mortonson}, M. 2003,
  \aj, 126, 1836

\bibitem[{{Hunter} {et~al.}(2000){Hunter}, {O'Connell}, {Gallagher}, \&
  {Smecker-Hane}}]{2000AJ....120.2383H}
{Hunter}, D.~A., {O'Connell}, R.~W., {Gallagher}, J.~S., \& {Smecker-Hane},
  T.~A. 2000, \aj, 120, 2383

\bibitem[{{Kharchenko} {et~al.}(2005){Kharchenko}, {Piskunov}, {R{\"o}ser},
  {Schilbach}, \& {Scholz}}]{2005A&A...438.1163K}
{Kharchenko}, N.~V., {Piskunov}, A.~E., {R{\"o}ser}, S., {Schilbach}, E., \&
  {Scholz}, R.-D. 2005, \aap, 438, 1163

\bibitem[{{Kroupa} \& {Boily}(2002)}]{2002MNRAS.336.1188K}
{Kroupa}, P. \& {Boily}, C.~M. 2002, \mnras, 336, 1188

\bibitem[{{Lada} \& {Lada}(2003)}]{2003ARA&A..41...57L}
{Lada}, C.~J. \& {Lada}, E.~A. 2003, \araa, 41, 57

\bibitem[{{Lamers} {et~al.}(2005{\natexlab{a}}){Lamers}, {Gieles}, {Bastian},
  {Baumgardt}, {Kharchenko}, \& {Portegies Zwart}}]{2005A&A...441..117L}
{Lamers}, H.~J.~G.~L.~M., {Gieles}, M., {Bastian}, N., {et~al.}
  2005{\natexlab{a}}, \aap, 441, 117

\bibitem[{{Lamers} {et~al.}(2005{\natexlab{b}}){Lamers}, {Gieles}, \&
  {Portegies Zwart}}]{2005A&A...429..173L}
{Lamers}, H.~J.~G.~L.~M., {Gieles}, M., \& {Portegies Zwart}, S.~F.
  2005{\natexlab{b}}, \aap, 429, 173

\bibitem[{{Larsen}(2000)}]{2000MNRAS.319..893L}
{Larsen}, S.~S. 2000, \mnras, 319, 893

\bibitem[{{Larsen}(2002)}]{2002AJ....124.1393L}
{Larsen}, S.~S. 2002, \aj, 124, 1393

\bibitem[{{Larsen} \& {Richtler}(1999)}]{1999A&A...345...59L}
{Larsen}, S.~S. \& {Richtler}, T. 1999, \aap, 345, 59

\bibitem[{{Larsen} \& {Richtler}(2000)}]{2000A&A...354..836L}
{Larsen}, S.~S. \& {Richtler}, T. 2000, \aap, 354, 836

\bibitem[{{Leitherer} {et~al.}(1999){Leitherer}, {Schaerer}, {Goldader},
  {Delgado}, {Robert}, {Kune}, {de Mello}, {Devost}, \&
  {Heckman}}]{1999ApJS..123....3L}
{Leitherer}, C., {Schaerer}, D., {Goldader}, J.~D., {et~al.} 1999, \apjs, 123,
  3

\bibitem[{{Maraston} {et~al.}(2004){Maraston}, {Bastian}, {Saglia},
  {Kissler-Patig}, {Schweizer}, \& {Goudfrooij}}]{2004A&A...416..467M}
{Maraston}, C., {Bastian}, N., {Saglia}, R.~P., {et~al.} 2004, \aap, 416, 467

\bibitem[{{Massey}(2002)}]{2002ApJS..141...81M}
{Massey}, P. 2002, \apjs, 141, 81

\bibitem[{{Mengel} {et~al.}(2005){Mengel}, {Lehnert}, {Thatte}, \&
  {Genzel}}]{mengel05}
{Mengel}, S., {Lehnert}, M.~D., {Thatte}, N., \& {Genzel}, R. 2005, ArXiv
  Astrophysics e-prints

\bibitem[{{Meurer} {et~al.}(1995){Meurer}, {Heckman}, {Leitherer}, {Kinney},
  {Robert}, \& {Garnett}}]{1995AJ....110.2665M}
{Meurer}, G.~R., {Heckman}, T.~M., {Leitherer}, C., {et~al.} 1995, \aj, 110,
  2665

\bibitem[{{Miller} {et~al.}(1997){Miller}, {Whitmore}, {Schweizer}, \&
  {Fall}}]{1997AJ....114.2381M}
{Miller}, B.~W., {Whitmore}, B.~C., {Schweizer}, F., \& {Fall}, S.~M. 1997,
  \aj, 114, 2381

\bibitem[{{Salo} \& {Laurikainen}(2000)}]{2000MNRAS.319..377S}
{Salo}, H. \& {Laurikainen}, E. 2000, \mnras, 319, 377

\bibitem[{{Schechter}(1976)}]{1976ApJ...203..297S}
{Schechter}, P. 1976, \apj, 203, 297

\bibitem[{{Schlegel} {et~al.}(1998){Schlegel}, {Finkbeiner}, \&
  {Davis}}]{1998ApJ...500..525S}
{Schlegel}, D.~J., {Finkbeiner}, D.~P., \& {Davis}, M. 1998, \apj, 500, 525

\bibitem[{{Schulz} {et~al.}(2002){Schulz}, {Fritze-v.~Alvensleben}, {M{\"
  o}ller}, \& {Fricke}}]{2002A&A...392....1S}
{Schulz}, J., {Fritze-v.~Alvensleben}, U., {M{\" o}ller}, C.~S., \& {Fricke},
  K.~J. 2002, \aap, 392, 1

\bibitem[{{Schweizer}(2005)}]{2005sdlb.proc..143S}
{Schweizer}, F. 2005, in ASSL Vol. 329: Starbursts: From 30 Doradus to Lyman
  Break Galaxies, 143--+

\bibitem[{{Solomon} {et~al.}(1987){Solomon}, {Rivolo}, {Barrett}, \&
  {Yahil}}]{1987ApJ...319..730S}
{Solomon}, P.~M., {Rivolo}, A.~R., {Barrett}, J., \& {Yahil}, A. 1987, \apj,
  319, 730

\bibitem[{{Weidner} {et~al.}(2004){Weidner}, {Kroupa}, \&
  {Larsen}}]{2004MNRAS.350.1503W}
{Weidner}, C., {Kroupa}, P., \& {Larsen}, S.~S. 2004, \mnras, 350, 1503

\bibitem[{{Whitmore}(2003)}]{2003dhst.symp..153W}
{Whitmore}, B.~C. 2003, in A Decade of Hubble Space Telescope Science, 153--178

\bibitem[{{Whitmore} \& {Schweizer}(1995)}]{1995AJ....109..960W}
{Whitmore}, B.~C. \& {Schweizer}, F. 1995, \aj, 109, 960

\bibitem[{{Whitmore} {et~al.}(1993){Whitmore}, {Schweizer}, {Leitherer},
  {Borne}, \& {Robert}}]{1993AJ....106.1354W}
{Whitmore}, B.~C., {Schweizer}, F., {Leitherer}, C., {Borne}, K., \& {Robert},
  C. 1993, \aj, 106, 1354

\bibitem[{{Whitmore} {et~al.}(1999){Whitmore}, {Zhang}, {Leitherer}, {Fall},
  {Schweizer}, \& {Miller}}]{1999AJ....118.1551W}
{Whitmore}, B.~C., {Zhang}, Q., {Leitherer}, C., {et~al.} 1999, \aj, 118, 1551

\bibitem[{{Williams} \& {McKee}(1997)}]{1997ApJ...476..166W}
{Williams}, J.~P. \& {McKee}, C.~F. 1997, \apj, 476, 166

\bibitem[{{Wilson} {et~al.}(2003){Wilson}, {Scoville}, {Madden}, \&
  {Charmandaris}}]{2003ApJ...599.1049W}
{Wilson}, C.~D., {Scoville}, N., {Madden}, S.~C., \& {Charmandaris}, V. 2003,
  \apj, 599, 1049

\bibitem[{{Zhang} \& {Fall}(1999)}]{1999ApJ...527L..81Z}
{Zhang}, Q. \& {Fall}, S.~M. 1999, \apjl, 527, L81

\end{thebibliography}

\end{document}